\documentclass[lettersize,journal]{IEEEtran}
\usepackage{amsmath,amsfonts}
\usepackage{upgreek}
\usepackage{bm}
\usepackage{algorithm} 
\usepackage{algorithmic}
\usepackage{array}
\usepackage{textcomp}
\usepackage{tabularx}
\usepackage{stfloats}
\usepackage{url}
\usepackage{verbatim}
\usepackage{caption}
\usepackage{romannum}
\usepackage{cite}
\usepackage{amsmath}
\usepackage{makecell}
\usepackage{multirow}
\usepackage{xcolor}
\usepackage{hyperref}
\usepackage{graphicx}
\usepackage{subcaption}
\DeclareRobustCommand\blue{\textcolor{blue}}
\newcolumntype{P}[1]{>{\centering\arraybackslash}p{#1}}
\begin{document}

\title{Network Interdependency-Informed Power System Dynamic Trajectory Prediction Utilizing Black-Box Modeling of Multiple Inverter-Based Resources}
\author{Sungjoo Chung,~\IEEEmembership{Graduate Student Member,~IEEE,}
        Ying Zhang,~\IEEEmembership{Member,~IEEE},
        Meng Yue, 
        Hantao Cui, ~\IEEEmembership{Senior Member,~IEEE}
\vspace{-25pt}
\thanks{S. Chung and Y. Zhang are with the School of Electrical and Computer Engineering, Oklahoma State University, Stillwater, OK 74074, USA (e-mail: sungjoo.chung@okstate.edu; y.zhang@okstate.edu). 

M. Yue is with Brookhaven National Laboratory, Upton, NY 11973, USA. 

H. Cui is with the Department of Electrical and Computer Engineering, North Carolina State University, Raleigh, NC 27695, USA.}}

\markboth{}%
{Shell \MakeLowercase{\textit{et al.}}: A Sample Article Using IEEEtran.cls for IEEE Journals}

\maketitle

\begin{abstract}
Black-box modeling of inverter-based resources (IBRs) has \textcolor{blue}{attracted growing interest} for real-time grid operation and control, especially when detailed proprietary electronic control architectures are inaccessible. Existing machine learning (ML)-based online dynamic trajectory prediction approaches using IBR black-box models either significantly accumulate prediction errors when multiple surrogates are simultaneously used or ignore measurement errors, limiting their deployment in practical grids. To address these limitations, this paper proposes a novel network interdependency-informed ML algorithm for online dynamic trajectory prediction in IBR-integrated power systems. A modular spatiotemporal attention network (STAN)-based predictor for the black-box modeling of each IBR unit is first proposed. Utilizing past measurements, the proposed STAN can effectively capture and predict the spatiotemporal dynamics of IBRs by employing an attention mechanism to attend to the most pertinent features for trajectory prediction. Furthermore, a novel hybrid physics-informed loss function that integrates a decoupled linearized AC power flow formulation is proposed.  The proposed loss function effectively ensures physical consistency of predictions within network operation while avoiding the computational complexity of iterative power flow solving, thereby enabling efficient gradient backpropagation and overall improved prediction accuracy. Case studies on the IEEE 14- and WECC 179-bus systems demonstrate that the proposed method achieves significant accuracy enhancement and robustness against measurement errors, outperforming recent ML-based methods.
\end{abstract}

\begin{IEEEkeywords} Dynamic trajectory prediction, inverter-based resources, black-box modeling, physics-informed neural networks, machine learning, data-driven.
\end{IEEEkeywords}
\vspace{-10pt}
\section{Introduction}

\IEEEPARstart{\blue{W}}{\blue{ith }}\blue{ advanced control strategies, inverter-based resources (IBRs) can provide fast voltage regulation, self-synchronization capability, and reliable operational support \cite{GFMControlRosso9408354}.} 
Grid-forming inverters have garnered growing attention due to their ability to regulate voltage, enabling fast ancillary services, such as reactive power support, to enhance grid resilience during contingencies \cite{DOE_ResilientGrid_7091092}.  \textcolor{blue}{However, as IBR penetration increases, the displacement of synchronous generators can reduce aggregate synchronous inertia and alter fault-current characteristics, making post-disturbance dynamics more difficult to predict.  In addition to these system-level dynamic challenges,} the firmware-based nature of IBRs is another issue, as original equipment manufacturers (OEMs) often restrict access to detailed control design due to confidentiality requirements. This limits the feasibility of conventional numerical and model-based approaches for dynamic studies, and thus, black-box modeling approaches for IBRs have attracted interest from academia and industry \cite{ DDModelingFan9761159}. This paper thus focuses on developing and utilizing black-box IBR modeling to construct an efficient dynamic trajectory prediction method in IBR-rich power systems. 
Such predictive models provide essential references for various real-time grid applications, such as online stability assessment \cite{Gupta8486644} and control \cite{zhang2024multi, Arjomandi10443528}. 

Conventional model-based approaches to dynamic trajectory simulation or assessment primarily rely on time-domain simulation \cite{EfficientTSA8372624, DecomposeTDS6923500} that solve nonlinear differential-algebraic equations (DAEs). While accurate, their reliance on detailed models creates high computational demand and makes them unsuitable for online deployment. 
Meanwhile, the proliferation of phasor measurement units (PMUs) has made high-resolution data available for machine learning (ML) techniques. \textcolor{blue}{In parallel with physics-based IBR modeling and offline validation (e.g., IEEE Std. 2800 \cite{IEEE2800-9762253}), advanced ML-based modeling can be learned from available measurement data and serve as a data-driven surrogate for the time-consuming DAE-solving procedure to improve computational efficiency}\cite{Ventura_Nadal_2025}. 

ML-based methods for dynamic trajectory prediction can be mainly divided into 1) data-driven ML \cite{JiamingMLOnlineTSA9302975, UseofMLYe10439652,StructureInformed9720101, FreqTrans10078020, BayesianTan10854889, TwinMahapatra10905168}, 2) physics-informed ML, which incorporates physical constraints and domain knowledge into the learning algorithm \cite{IntegratingLiJiaming10066348, OnlineDSAHybrid10632211, Ventura_Nadal_2025, PhysicsInformedPTSA10542429, GrayZhang10507020}. 
In \cite{JiamingMLOnlineTSA9302975}, a long short-term memory (LSTM)-based recurrent neural network is constructed for online prediction of the dynamic state of a synchronous generator (SG), where voltage measurements from neighboring buses are used as additional inputs. A one-shot LSTM-based method for the prediction of post-fault rotor angle trajectories considering PMU noise and time delays is developed in \cite{UseofMLYe10439652}. Yet, both methods focus on predicting single-component-level dynamics and do not consider the network coupling, which can compromise systematic prediction accuracy. To address this limitation, \cite{StructureInformed9720101} introduces a similarity matrix-based graph learning approach to capture correlations between SGs, enabling topology-informed prediction of transient dynamics. Nonetheless, this method is susceptible to error accumulation stemming from its reliance on rolling prediction schemes. To reduce the accumulating error and high computational burden from rolling predictions in the time domain, \cite{FreqTrans10078020} develops a neural network that learns the spatiotemporal transient dynamics in the frequency domain. A two-stage detection and trajectory prediction mechanism is developed in \cite {BayesianTan10854889}, which first identifies in-synchronization and out-of-step generators using a data-driven detector. Recently, a digital twin framework using PMU data for IBR black-box modeling was introduced in \cite{TwinMahapatra10905168}, but it focuses on the grid edge and neglects the PMU errors.

To improve consistency with physical laws, physics-informed ML approaches have been explored by embedding power flow formulation \cite{IntegratingLiJiaming10066348}, governing/swing equations \cite{OnlineDSAHybrid10632211, Ventura_Nadal_2025}, or network topology \cite{PhysicsInformedPTSA10542429, GrayZhang10507020}. For instance, \cite{IntegratingLiJiaming10066348} \textcolor{blue}{considers the full network information} and adopts AC power flow (ACPF)-computed voltages at the neighboring buses of an SG bus as inputs to enhance learning effectiveness. However, the scalability of this approach to a large-scale system with multiple SGs remains an issue. Two physics-informed neural networks (PINNs) embed the swing equation into their loss functions in \cite{OnlineDSAHybrid10632211} to model the system dynamics under transient stable and unstable conditions. However, PMU measurement errors are neglected. A plug-and-play PINN, which acts as a surrogate model at the component level, is integrated into time-domain system simulation in \cite{Ventura_Nadal_2025}. Compared with most existing research on predicting the transient responses of SGs \cite{IntegratingLiJiaming10066348, OnlineDSAHybrid10632211, Ventura_Nadal_2025, PhysicsInformedPTSA10542429}, the literature on IBR trajectory prediction is limited. A Kron-reduction-based optimization is integrated with a deep neural network in \cite{GrayZhang10507020}, but the IBR modeling method is a white-box approach and requires complete knowledge of the internal control details. 

Existing ML-based dynamic trajectory prediction methods have several limitations, and only a handful of papers \cite{GrayZhang10507020, TwinMahapatra10905168} are designed for IBR-integrated power systems. First, most of these methods focus on a single dynamic component, either IBR or SG, and \textcolor{blue}{are separately trained based on individual neural network design. When multiple surrogates are deployed simultaneously in the grid}, neglecting their network interdependencies can lead to a significant accumulation of prediction errors. Second, these methods often overlook PMU measurement errors in practical environments, and their limited robustness to such errors further amplifies error propagation throughout network operation. Lastly, IBRs exhibit fast dynamics and require real-time monitoring and control, and existing methods developed for SGs may be unsuitable for IBRs \cite{Arjomandi10443528}. These gaps motivate the pursuit of an efficient solution to dynamic trajectory prediction of multiple grid-tied IBRs, ensuring systematic accuracy in the presence of measurement errors.

To fill the gaps, this paper proposes a network interdependency-informed ML algorithm for online prediction of power system dynamics using black-box IBR models. A modular spatiotemporal attention network (STAN)-based predictor for black-box modeling of each IBR unit is first developed. It leverages the temporal dynamics of IBR buses and the spatial correlations between the dynamic behavior of neighboring buses. The attention mechanism enables the model to selectively learn the pertinent spatiotemporal dynamics, effectively improving its prediction accuracy. 
\textcolor{blue}{Furthermore, considering the full power transmission network, a decoupled linearized ACPF (DL-ACPF) formulation is adopted to form a hybrid physics-informed loss function. This formulation avoids the iterative update of the Jacobian matrix by a linearized decoupled approximation of the network model, yielding an analytical power flow solution. The hybrid physics-informed loss function with this analytical expression} regularizes the voltage predictions of all grid-tied IBR surrogates toward coordinated and physically consistent trajectories while enabling efficient gradient computation.

The main contributions are summarized as follows:

\begin{itemize}
    \item \textcolor{blue}{Unlike existing ML-based methods that are separately trained for individual IBR or SG \cite{TwinMahapatra10905168,IntegratingLiJiaming10066348}, the proposed method jointly trains all the IBR surrogate models by embedding system-wide network operation into the learning process, significantly enhancing their global accuracy.} 
    \item The proposed algorithm can simultaneously capture temporal dependencies in past timesteps and spatial network interdependencies. This produces an enhanced IBR modeling surrogate that requires no knowledge of proprietary control structures.
    \item The proposed hybrid physics-informed loss function enables coordinated prediction of IBR surrogates, ensuring physical consistency of dynamic trajectory predictions across the grid. This effectively mitigates the error accumulation inherent in rolling prediction and the performance degradation caused by measurement errors. 
    \item Comparative studies with recent data-driven and physics-informed ML methods confirm the proposed method’s effectiveness in fast, accurate, and physics-consistent IBR dynamic prediction for grid operation. Utilizing the proposed algorithm, the dynamic trajectories for all non-metered buses can be predicted across large-scale power systems efficiently. 
\end{itemize}

\textcolor{blue}{\textit{Notation}: In the paper, scalars are denoted by non-bold lowercase, vectors by bold lowercase, and matrices by bold uppercase; the symbols $\mathbb{C}$ and $\mathbb{R}$ denote the fields of complex and real numbers, respectively. 
Unless otherwise specified, variables are real-valued. }

\section{Preliminaries and Problem Formulation} \label{Preliminaries and Problem Formulation}

The dynamics of power grids are governed by a set of nonlinear DAEs. Without loss of generality, the DAEs can be expressed as:
\begin{align} 
     & \boldsymbol{\dot{x}}(t) = \mathbf{f}(\mathbf{x}(t),\mathbf{z}(t)) \label{DE}  \\
     & 0 = \mathbf{g}({\mathbf{x}(t),\mathbf{z}}(t))  \label{AE}
\end{align}
where $\mathbf{x}$ denotes the dynamic state vector, and $\mathbf{z}$ denotes the algebraic state vector;  \eqref{DE} represents the differential equations (DEs) of the dynamic components in the network, and \eqref{AE} represents the algebraic equations (AEs) corresponding to the network equations. Note that the dynamics of IBRs are coupled through the network connections in \eqref{AE}.

The dynamics of a power system are governed by several factors, such as IBR control structures, generator dynamics, dynamic load characteristics, and network topology. While the internal states of SGs are well-defined, the proprietary nature of inverter controls renders the conventional numeric iterative approach to solve \eqref{DE} and \eqref{AE} impractical \cite{FanLingling9269455}.

\subsection{Control Model of Grid-Forming Inverters }

\begin{figure}
    \centering
    \includegraphics[width=0.7\linewidth]{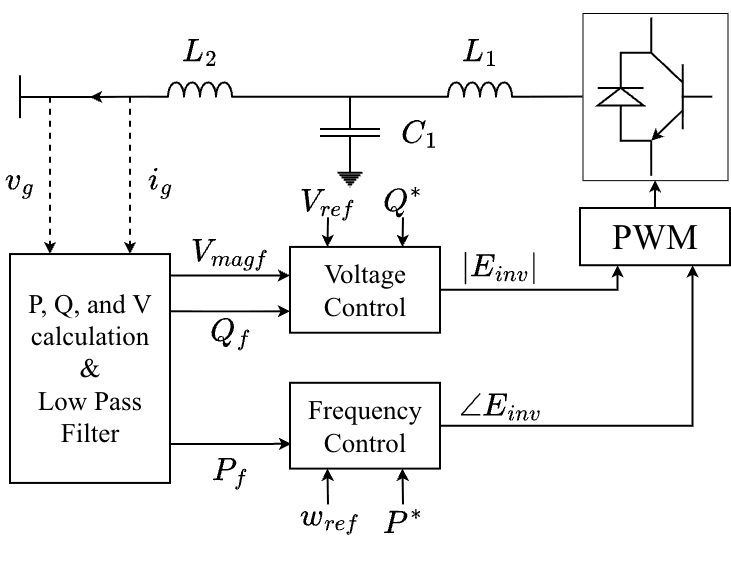}
     \vspace{-10pt}
    \caption{Control structure of a \textcolor{blue}{droop-controlled} grid-forming inverter.}
    \label{fig:GFM_General}
        \vspace{-10pt}
\end{figure}


A grid-forming inverter behaves approximately as a voltage source at its connected bus, controlling voltage and frequency to balance loads and provide reactive power support\cite{DOE_ResilientGrid_7091092}. \textcolor{blue}{As a proof-of-concept paper, the proposed method focuses on droop control, which is widely adopted for system-wide frequency synchronization and power sharing among grid-forming inverters. The proposed method can be extended to other inverter control types, for example, by data-driven model switching detection \cite{HuangHeqing11313680} and then grouping the training dataset; further investigation will be left for our future work.} 

Fig. \ref{fig:GFM_General} depicts the control structure of \textcolor{blue}{a droop-controlled} grid-forming inverter. The bus voltage and current, denoted as $v_g$ and $i_g$, are used to compute the instantaneous active power $P$, reactive power $Q$, and voltage magnitude $V_{mag}$ in the $\alpha \beta$ reference frame, which are filtered using first-order low-pass filters to yield $P_f$, $Q_f$, and $V_{magf}$. The voltage and frequency references, $V_{ref}$ and $\omega_{ref}$, along with the active and reactive power setpoints, $P^*$ and $Q^*$, are combined with the filtered signals to compute the voltage magnitude $|E_{inv}|$ and phase angle $\angle E_{inv}$, depending on the control strategy. Finally, the inverter output is synthesized via pulse-width modulation (PWM) based on the computed voltage phasor.
The droop control laws determining the inverter voltage and frequency setpoints in Fig. \ref{fig:GFM_General} are given by
\begin{align}
      V& =V_{ref}-m_{q}(Q_{f}-Q^{\ast})\\ 
    \omega& =\omega_{ref}-m_{p}(P_{\text{f}}-P^{\ast}) 
\end{align}
where $m_{p}$ and $m_{q}$ are the droop coefficients for active and reactive power control, respectively; $V$ and $\omega$ as intermediate control variables are subsequently processed to derive the terminal voltage phasor $e_{inv}$. 
A more detailed description of the droop control adopted in this paper can be found in \cite{Du9173757}.
\vspace{-8 pt}
\subsection{Problem Formulation of Black-Box IBR Modeling} \label{Problem Formulation}

Given sudden contingencies such as generator loss or load shedding in the grid, enhanced black-box modeling of IBR nodes can serve as an efficient surrogate to quickly determine the voltage status of all IBR nodes across the grid. 
The proposed black-box model leverages historical sequences of PMU measurements to predict the voltages of IBR nodes, enabling fast and accurate online inference for various operating conditions. 

Assume that a limited number of PMUs are installed only on key nodes, such as those with IBRs. This assumption aligns with the current research and engineering practice \cite{Guo7169632, Gupta8486644, TwinMahapatra10905168}.
\textcolor{blue}{Capturing sub-synchronous oscillations and other higher-frequency phenomena will require electromagnetic transient (EMT) simulation and measurements, which are out of the scope of this paper.} 
By utilizing these historical PMU voltage data and related grid operating records (i.e., DER generation and load consumption), the mapping inside the IBRs, which is conventionally characterized by DEs, can be learned offline. This learned mapping can then be utilized for online inference to quickly and accurately predict the dynamic response of IBRs under various contingencies.
\begin{figure}
    \centering
    \includegraphics[width=\columnwidth]{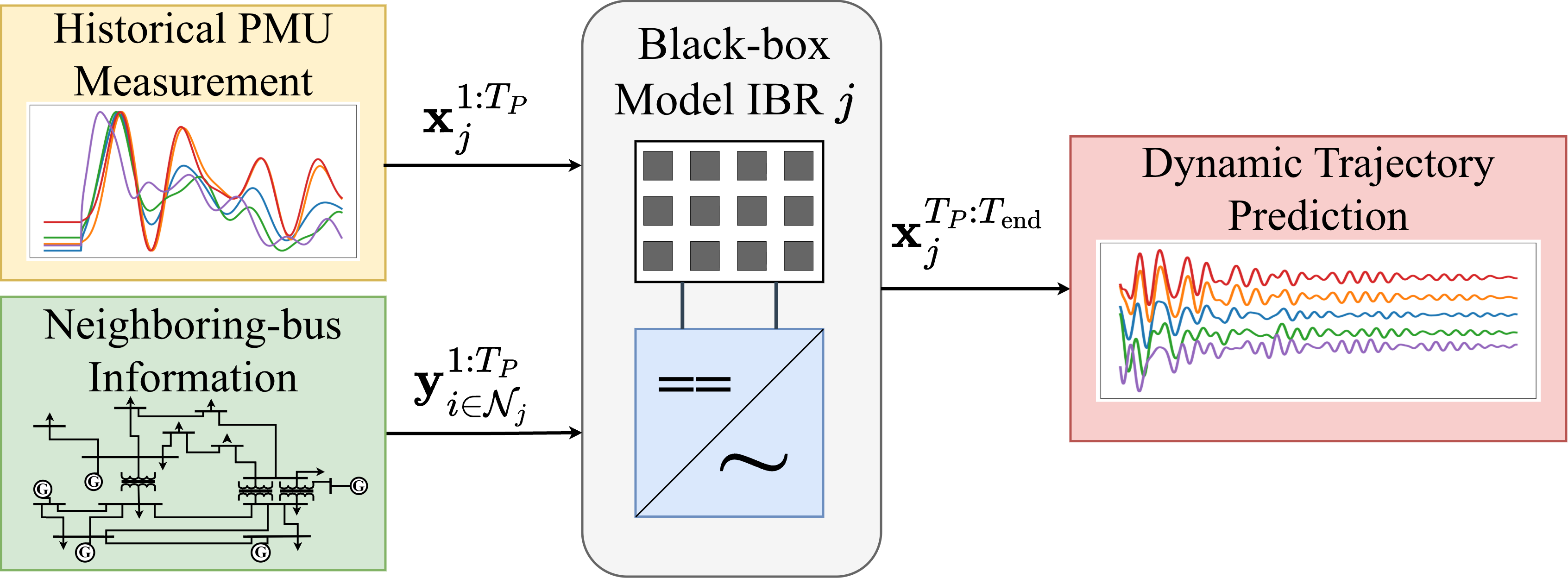}
    \caption{Schematic diagrams of black-box modeling of a single inverter-based resource.}
    \label{fig:BBox}
\end{figure}
The black-box modeling of IBR bus $j$, under different grid operating records, is illustrated in Fig. \ref{fig:BBox}. Given $T_P$ historical timesteps of the voltage of IBR bus $j$, denoted as $\mathbf{x}_{j}^{1:T_{P}} = [\mathbf{x}^{1}_{j},\dots,\mathbf{x}_{j}^{T_P}]$, where  $\mathbf{x}_{j}^{t}  = [ |v_{j}^{t}|,\: \angle v_{j}^{t}]^\top$, the black-box model predicts the next-step voltage state as
\begin{align}\label{Bbox_onestep}
    \hat{\mathbf{x}}^{T_{P}+1}_j = \mathcal{F}(\mathbf{x}_j^{1:T_{P}}, \mathbf{y}_{i\in \mathcal{N}_j}^{1:T_{P}})
\end{align}
where  $\mathbf{y}_{i\in \mathcal{N}_j}^{}$ denotes external information from the power system defined as the voltages from the neighboring set of all nodes connected to bus $j$, denoted as $\mathcal{N}_j$. \textcolor{blue}{In the remainder of this paper, ``voltage'' refers to the voltage magnitude and phase angle, unless otherwise specified.}

Note that voltages of IBR buses are treated as dynamic variables in \eqref{DE}, as they are actively governed by their inverter control. Moreover, voltages of IBR buses are selected as the prediction target because they are externally observable from PMU measurements and directly support bus-level dynamic trajectory assessment. Once the voltage magnitude and phase angle are obtained, IBR active and reactive power injections can be further estimated through the network equations. In cases where PMU installation is unavailable to provide high-resolution voltage data, simulated data obtained by solving the DAEs through conventional methods is considered as a supplement. 

The single-step prediction method in \eqref{Bbox_onestep} can be extended to a multi-step prediction method \cite{StructureInformed9720101} by
\begin{align} \label{Bbox_multstep}
    \hat{\mathbf{x}}^{T_{P}+m}_j = \mathcal{F}(\mathbf{x}_j^{1:T_{P}}, \mathbf{y}_{i\in \mathcal{N}_j}^{1:T_{P}}) \quad m=1,\dots, T_{R}
\end{align}
where $T_{R}$ is the number of prediction steps. To obtain the trajectory up to the final timestep $T_{\text{end}}$, this multi-step predictor is applied recursively. In \eqref{Bbox_onestep} and \eqref{Bbox_multstep}, only the initial predictions use measured states $\mathbf{x}^{1:T_P}_j$ and $\mathbf{y}^{1:T_P}_{i\in \mathcal{N}_j}$, while subsequent steps rely on prior predictions of the black-box model.

During online inference, the black-box models, trained with local PMU measurements and external network information, predict the voltage state vector $\hat{\mathbf{x}}^{t}$ for all IBR buses at each timestep, serving as a fast surrogate for solving the DEs in \eqref{DE}. Note that incorporating external information from the network can enhance the prediction accuracy of black-box modeling \cite {IntegratingLiJiaming10066348}. Therefore, we propose a novel physics-informed algorithm to inform each IBR surrogate model of the power network operation, which is detailed in the next section.

\begin{figure}
    \centering
    \includegraphics[width=0.9\linewidth]{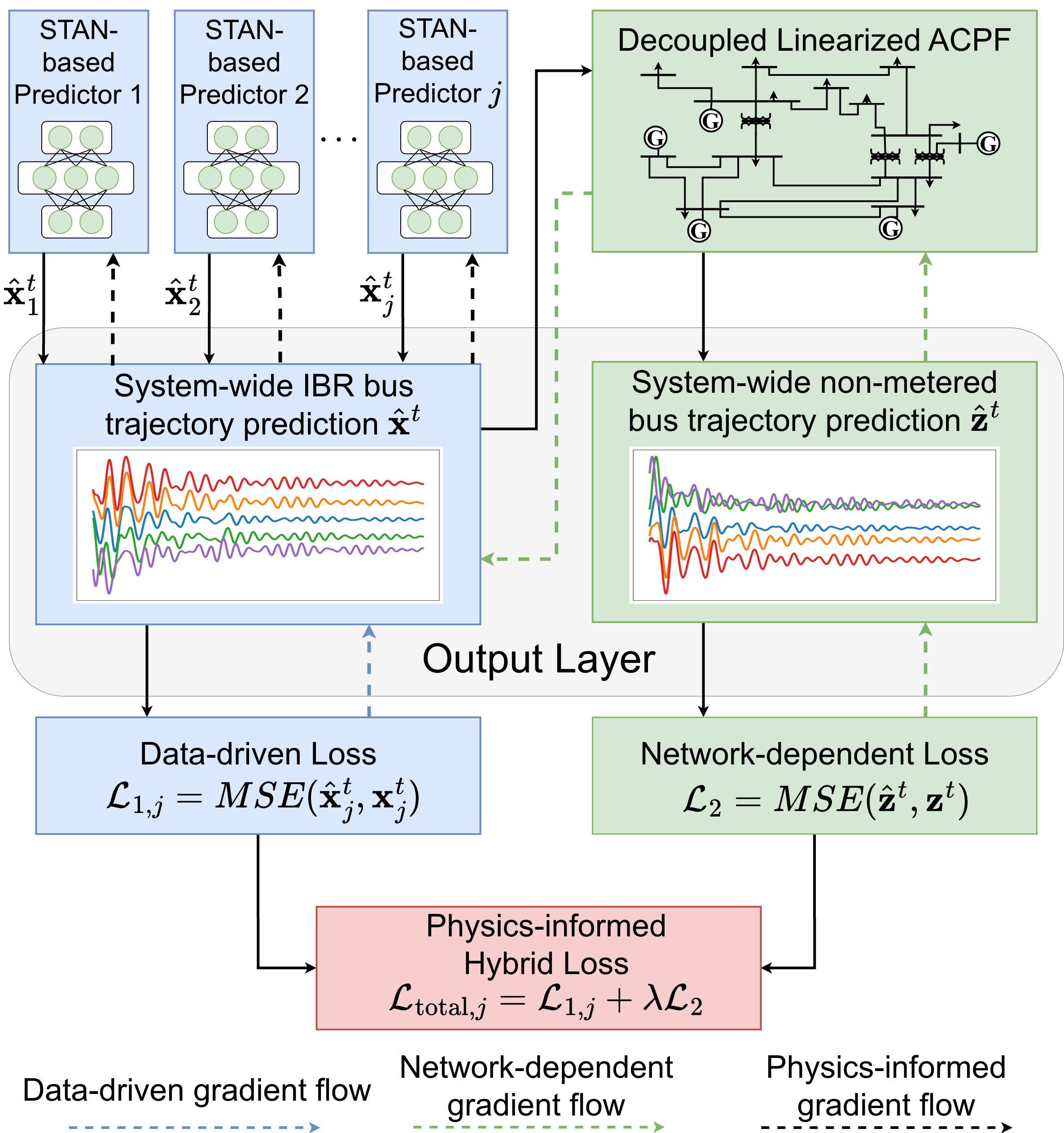}
    \caption{Overall architecture of the proposed network interdependency-informed framework. \textcolor{blue}{The information exchange is denoted as solid arrows in the forward pass and dashed arrows in the back pass.} }
    \label{fig:Architecture}
\end{figure}

\section{Proposed Network Interdependency-Informed Dynamic Trajectory Prediction Method}

A network interdependency-informed STAN-based framework for online dynamic trajectory prediction is proposed. Illustrated in Fig. \ref{fig:Architecture}, with historical PMU measurements and estimated neighboring bus voltages as the input, each STAN-based predictor predicts the dynamics of that IBR unit by capturing the spatiotemporal dependencies.  To further enhance the overall prediction accuracy, a novel physics-informed hybrid loss function is implemented to ensure physical consistency for the predicted dynamic trajectories of all the IBR buses. Moreover, this modularity allows the proposed method to be easily scaled to predict the dynamics of IBR and non-metered buses in large-scale power systems. These predictions are then used in an analytical DL-ACPF calculation to compute system-wide dynamic trajectories for all IBR and non-metered buses efficiently. 

\vspace{-10pt}
\subsection{Spatiotemporal Attention Network-Based Predictor} \label{sec: STAN}

\begin{figure}
    \centering
    \includegraphics[width=0.82\linewidth]{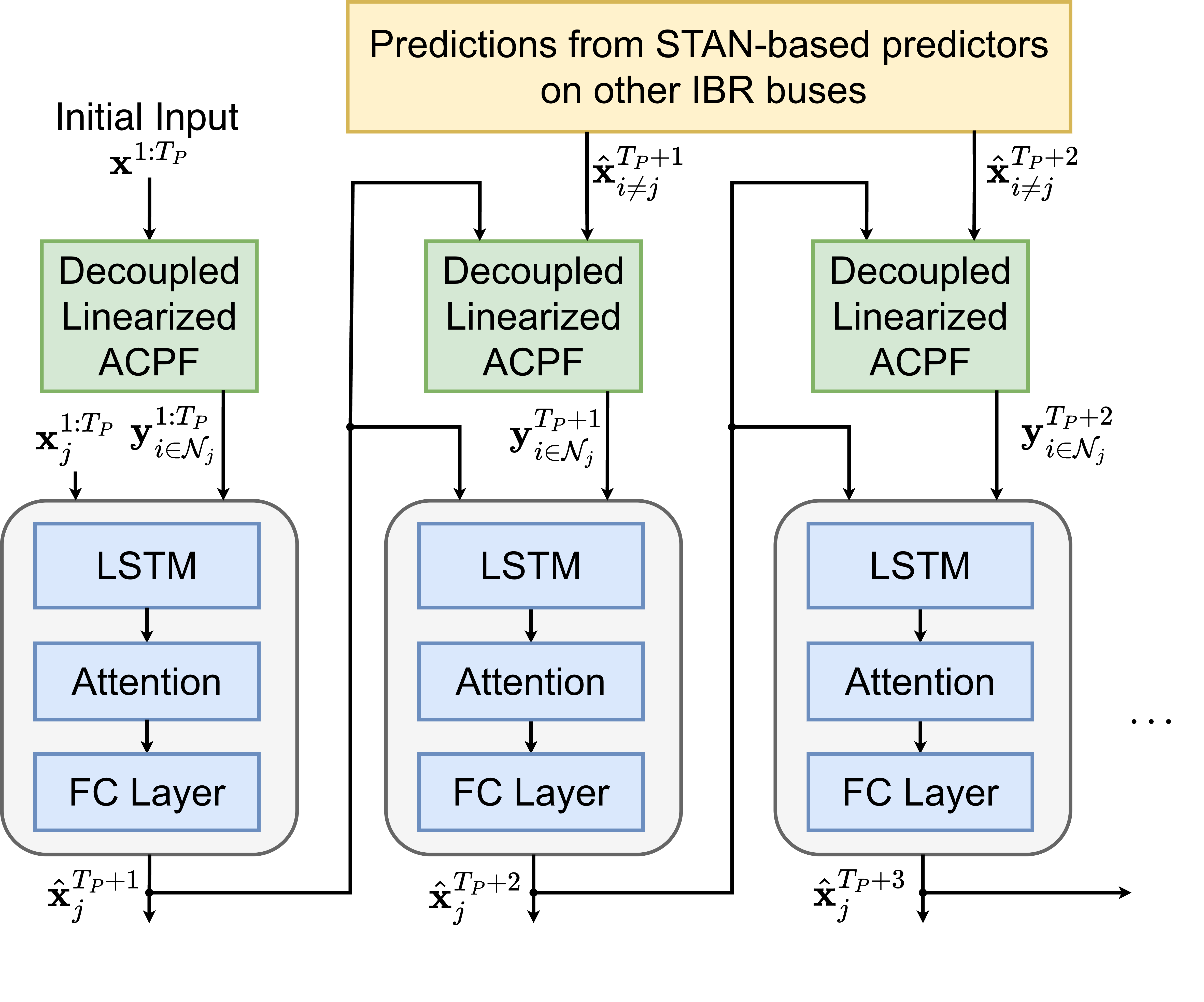}
    \caption{Proposed STAN-based predictor architecture for IBR bus $j$.}
    \label{fig:Predictor_arch}
\end{figure}

The architecture of an individual STAN-based predictor is illustrated in Fig. \ref{fig:Predictor_arch}. This predictor serves as a black-box model for individual IBR buses, predicting next-step voltage based on historical PMU measurements. To further enhance prediction accuracy and robustness, the voltages of neighboring non-IBR buses are incorporated as additional network information and denoted as  $\mathbf{y}^{1:T_P}_{i\in\mathcal{N}_{j}}$, \textcolor{blue}{where each neighboring-bus voltage state is represented as $\mathbf{y}^t_{i}=[\lvert v_{i}^t \rvert, \angle v_{i}^t ]^\top$}. This integration leverages the local spatiotemporal correlations in power systems, as neighboring bus voltage dynamics are physically coupled through network admittances.

The proposed STAN-based predictor employs a two-layer LSTM network to capture the temporal correlations in the voltage sequences of an IBR bus and its neighboring buses. Based on this initial input sequence, the hidden state of the black-box model of the IBR bus $j$ at time $t$ is obtained by
\begin{align} 
    \mathbf{h}_j^t &= lstm(\mathbf{x}_j^{1:T_{P}}, \mathbf{y}_{i\in \mathcal{N}_j}^{1:T_{P}}) \label{LSTM_hidden}
\end{align}
where the detailed LSTM operation $lstm(\cdot)$ is described in the Appendix.

While conventional recurrent neural networks can selectively retain information through their gating mechanism, they treat all past hidden states uniformly. However, not all past hidden states are equally informative for predicting the next-step voltages of IBR buses. To address this, an attention mechanism is employed, which uses query, key, and value vectors to dynamically evaluate the relevance of each feature. This allows the model to focus on the most pertinent features for prediction while attenuating the influence of less relevant ones. First, a query vector $\mathbf{q}_j$ is computed from the hidden state $\mathbf{h}_j(t)$ to determine the relevance of previous hidden states for next-step predictions:
\begin{align}
    \mathbf{q}_{j} = \mathbf{W}_{q} \mathbf{h}_j(t) + \mathbf{b}_{q}
\end{align}
where $\mathbf{W}_q$ and $\mathbf{b}_q$ are learnable parameters of the attention mechanism. 

Then, each hidden state $\mathbf{h}_{j}(\tau)$ from all the previous timesteps, $\tau=1, 2, \dots,t$, is transformed into a key-value pair:
\begin{align}
    \mathbf{k}_{j,\tau} &= \mathbf{W}_{k}\mathbf{h}_j(\tau) + \mathbf{b}_{k}\\
    \mathbf{s}_{j,\tau} &= \mathbf{W}_{s}\mathbf{h}_j(\tau) + \mathbf{b}_{s}  
\end{align}
where $\mathbf{W}_{k}$, $\mathbf{W}_{s}$, $\mathbf{b}_{k}$, and $\mathbf{b}_{s}$ are learnable parameters of the attention mechanism. 

Attention scores are calculated by the normalization of the dot product between the set of keys $\mathbf{k}_j = [\mathbf{k}_{j, 1}, \dots, \mathbf{k}_{j, t}]$ and the query:
\begin{align}
    \mathbf{a}_j=\text{softmax}\left(\frac{\mathbf{q}_j\mathbf{k}_j^{\top}}{\sqrt{d_k}}\right), \quad \mathbf{a}_j = [\mathbf{a}_{j, 1}, \dots, \mathbf{a}_{j, t}]
\end{align}
where $d_k$ is the dimension of the key vector $\mathbf{k}_j$.

The attention scores $\mathbf{a}_j$ then determine the contribution of each value vector, and their weighted sum produces the aggregated output, known as the context vector:
\begin{align}
\mathbf{c}_j = \sum_{\tau=1}^{t} \mathbf{a}_{j,\tau} \mathbf{s}_{j,\tau}
= \mathbf{a}_j\mathbf{s}_j
\end{align}
where the corresponding set of values $\mathbf{s}_j = [\mathbf{s}_{j, 1}, \dots, \mathbf{s}_{j,t}]$ represents the information in each hidden state. 

Finally, the output of the $j$th STAN is obtained by combining the context vector $\mathbf{c}_j$ with the current hidden state and passing it through a fully connected output layer:
\begin{align} \label{FC_layer}
\hat{\mathbf{x}}_j^{t+1} = \mathbf{W}_o \big[ \mathbf{h}_j(t); \mathbf{c}_j \big] + \mathbf{b}_o
\end{align}
where $\mathbf{W}_o$ and $\mathbf{b}_o$ are learnable parameters of the output layer, while $[\cdot; \cdot]$ denotes vector concatenation. This concatenation combines the attention-weighted historical information $\mathbf{c}_j$ with the current LSTM state $\mathbf{h}_j(t)$, enabling the predictor to leverage both attended past features and current temporal context for more accurate next-step predictions.

A data-driven loss function for each predictor is formulated based on its voltage predictions, denoted as $\hat{\mathbf{x}}^t_j$, and the \textcolor{blue}{ground-truth voltage obtained from historical records or time-domain simulations}, denoted as $\mathbf{x}^t_j$, by using the mean squared error (MSE):
\begin{align} \label{MSE}
    \mathcal{L}_{1, j}(\hat{\mathbf{x}}_{j}, \mathbf{x}_{j} ; {\Phi}_{j}) = \frac{1}{T_{\text{end}} - (T_{P}+1)}\sum_{t=T_{P}+1}^{T_{\text{end}}}\lVert\hat{\mathbf{x}}^{t}_{j} - \mathbf{x}^{t}_{j}\rVert^2
\end{align}
where ${\Phi}_j$ denotes the learnable parameters of the $j$-th STAN, ${\Phi}_j=\{\mathbf{W}_{\cdot, j},\mathbf{b}_{\cdot, j},...\}$. The backpropagation of the data-driven gradients updates the learnable parameters of the STAN such that it produces predictions to match the ground-truth voltages. The details of the gradient update procedure will be presented in Section \ref{Grad}.

In addition to the data-driven loss function \eqref{MSE}, network interdependency is incorporated into the loss function in the proposed method to enhance the prediction accuracy of the STAN-based predictors. This yields a hybrid physics-informed loss function that enforces system-wide physical consistency of the trajectory predictions, which will be introduced in Section \ref{sec: Physics-Informed}.

\subsection{Incorporating Network Interdependency} \label{sec: Physics-Informed}

Solely relying on the data-driven loss function in \eqref{MSE} may make the black-box model highly susceptible to measurement errors. This issue becomes especially pronounced when multiple ML-based predictors, locally trained on error-prone measurements, are collectively utilized for power system operations. Moreover, minimizing \eqref{MSE} independently for each IBR bus neglects the coupling among buses imposed by the underlying power flow equations. Consequently, the predictions may fit local trajectories well but violate global network interdependencies, leading to physically implausible voltage predictions. To overcome the limitations of the purely data-driven loss function, a hybrid physics-informed loss function is proposed, ensuring system-wide physical consistency of the predictions by incorporating power flow calculations.

\subsubsection{ Proposed Hybrid Physics-Informed Loss Function}

\textcolor{blue}{As illustrated in Figs. \ref{fig:Architecture} and \ref{fig:Predictor_arch}, the proposed STAN predictors exchange information with the network model by providing the predicted voltages at the IBR buses as inputs to the DL-ACPF formulation. Specifically,} the STAN predictions of all IBR buses at each timestep, denoted as $\hat{\mathbf{x}}^t = [\hat{\mathbf{x}}^t_1, \dots, \hat{\mathbf{x}}^t_{{N_{\text{IBR}}}}]$, are used to compute the voltage magnitudes and phase angles at the remaining network buses through power flow calculations. The resulting voltage magnitudes and phase angles constitute the STAN prediction-based power flow solution, denoted as $\hat{\mathbf{z}}^t$, which will be introduced later. Comparing $\hat{\mathbf{z}}^t$ with the target power flow solution $\mathbf{z}^t$ from historical records yields the network-dependent loss function:
\begin{align} \label{PF_MSE}
    \mathcal{L}_{2}(\mathbf{\hat{z}}, \mathbf{z} ; \Phi) = \frac{1}{T_{\text{end}}-(T_P+1)}\sum_{t=T_P+1}^{T_{\text{end}}} \lVert\mathbf{\hat{z}}^t - \mathbf{z}^t\rVert^2
\end{align}
where $\Phi=\{\Phi_1, \dots, \Phi_{N_{\text{IBR}}} \}$ represents the learnable parameters of all STANs in the power network; $N_{\text{IBR}}$ denotes the total number of IBR buses.

The final loss function used in the proposed method is a weighted sum of \eqref{MSE}  and \eqref{PF_MSE} to simultaneously take spatiotemporal correlations and global network dependencies into account, expressed as
\begin{align} \label{MSE_Total}
    \mathcal{L}_{\text{total}, j} = \mathcal{L}_{1, j} + \lambda \mathcal{L}_{2}
\end{align}
where $\lambda$ denotes the regularization coefficient that balances the contribution of the data-driven and network-dependent loss functions to trajectory predictions.

The proposed hybrid loss function in \eqref{MSE_Total} ensures that the predicted IBR voltages are physically consistent with the underlying power flow equations. Furthermore, because errors in any individual IBR prediction affect the accuracy of the overall power flow solution due to interdependencies among buses, this loss function encourages coordinated predictions that preserve the system-wide physical consistency of all STAN predictions.

\subsubsection{Calculating $\hat{\mathbf{z}}^t$ in the Proposed Loss Function \eqref{MSE_Total}}

\textcolor{blue}{A DL-ACPF formulation proposed in \cite{YangJingwei7782382} is adopted to construct $\mathcal{L}_{2}$. This formulation avoids the iterative update of the Jacobian matrix by providing a linear approximation of the network model. At each timestep in multi-step trajectory prediction, the proposed method computes the Jacobian matrix only once. Therefore, an analytical DL-ACPF solution $\hat{\mathbf{z}}^{t}$  can be computed directly at each timestep from the collective predictions of all STAN predictors. This direct computation enables efficient backpropagation of the physics-informed loss $\mathcal{L}_2$ obtained in \eqref{PF_MSE} through the predicted trajectory, while regularizing and coordinating all STAN predictions toward network physical consistency.}


Using the approximation techniques proposed in \cite{YangJingwei7782382}, the DL-ACPF model is expressed in compact matrix form as:
\begin{align} \label{DLPF_1}
    \begin{bmatrix} \mathbf{p} \\ \mathbf{q} \end{bmatrix} = - 
    \begin{bmatrix} \mathbf{B}^{\prime} & -\mathbf{G} \\ \mathbf{G} & \mathbf{B} \end{bmatrix} 
    \begin{bmatrix} \boldsymbol{\theta} \\ \mathbf{v} \end{bmatrix}
\end{align}
where $\mathbf{p}$ and $\mathbf{q}$ are active and reactive power injection vectors, and $\boldsymbol{\theta}$ and $\mathbf{v}$ are phase angle and voltage magnitude vectors, respectively; $\mathbf{G}$ and $\mathbf{B}$ are the conductance and susceptance matrices of the nodal admittance matrix $\boldsymbol{Y}=\mathbf{G}+j\mathbf{B}$; $\mathbf{B}^{\prime}$ denotes the susceptance matrix without shunt elements. \textcolor{blue}{The nodal admittance matrix in \eqref{Y_partition} characterizes the pre-fault network parameters and is assumed known to the system operator \cite{BayesianTan10854889}. The fault-induced effects are learned from post-disturbance trajectory data; thus, the proposed method does not require an up-to-date post-fault admittance matrix.} 


\textcolor{blue}{All the buses are divided into the reference angle-setting, voltage-controlled}, and PQ bus sets, denoted by $\mathcal{S}$, $\mathcal{R}$, and $\mathcal{W}$, respectively.
\textcolor{blue}{The IBR buses are included in the voltage-controlled bus set $\mathcal{R}$, and their voltages are predicted by the STAN at each timestep.} \textcolor{blue}{Subscripts $\mathcal{S}$, $\mathcal{R}$, and $\mathcal{W}$ denote the corresponding subvectors or submatrices associated with these bus sets.}
With this bus partitioning, $\boldsymbol{Y}$ is partitioned as the following block matrices:
\begin{align} \label{Y_partition}
    \boldsymbol{Y}=\left[\begin{array}{lll}
\boldsymbol{Y}_{\mathcal{S} \mathcal{S}} & \boldsymbol{Y}_{\mathcal{S} \mathcal{R}} & \boldsymbol{Y}_{\mathcal{S} \mathcal{W}} \\
\boldsymbol{Y}_{\mathcal{R} \mathcal{S}} & \boldsymbol{Y}_{\mathcal{R} \mathcal{R}} & \boldsymbol{Y}_{\mathcal{R} \mathcal{W}} \\
\boldsymbol{Y}_{\mathcal{W} \mathcal{S}} & \boldsymbol{Y}_{\mathcal{W} \mathcal{R}} & \boldsymbol{Y}_{\mathcal{W} \mathcal{W}}
\end{array}\right]
\end{align}
\textcolor{blue}{where
the dimension of each submatrix is determined by the size of the corresponding bus sets; for example, $\boldsymbol{Y}_{\mathcal{R}\mathcal{W}}\in\mathbb{C}^{|\mathcal{R}|\times |\mathcal{W}|}$.} 

The phase angle and voltage magnitude vectors to be solved in the partitioned formulation in \eqref{DLPF_1} are expressed as:
\begin{align} 
    {\tilde{\boldsymbol{\theta}}} &=
\begin{bmatrix}
    \boldsymbol{\theta}^\top_{\mathcal{R}}, 
    \boldsymbol{\theta}^\top_{\mathcal{W}}
\end{bmatrix}^\top \label{PF_sol1} \\
    {\tilde{\mathbf{v}}}& = \mathbf{v}_{\mathcal{W}} \label{PF_sol2}
\end{align}

With the arrangement in \eqref{Y_partition}-\eqref{PF_sol2}, \eqref{DLPF_1} is rearranged as
\begin{align} \label{DLPF_2}
    \begin{bmatrix} \tilde{\mathbf{p}} \\ \tilde{\mathbf{q}} \end{bmatrix} =  
    \begin{bmatrix} \mathbf{H} & \mathbf{N} \\ \mathbf{M} & \mathbf{L} \end{bmatrix} 
    \begin{bmatrix} \tilde{\boldsymbol{\theta}} \\ \tilde{\mathbf{v}} \end{bmatrix}
\end{align}
where $\mathbf{H}$, $\mathbf{N}$, $\mathbf{M}$, and $\mathbf{L}$ represent the corresponding conductance and susceptance matrices:
\begin{align}
    \left[\begin{array}{cc}
    \mathbf{H} & \mathbf{N} \\
    \mathbf{M} & \mathbf{L}
    \end{array}\right]=-\left[\begin{array}{cc|r}
    \mathbf{B}_{\mathcal{R} \mathcal{R}}^{\prime} & \mathbf{B}_{\mathcal{R} \mathcal{W}}^{\prime} & -\mathbf{G}_{\mathcal{R} \mathcal{W}} \\
    \mathbf{B}_{\mathcal{W} \mathcal{R}}^{\prime} & \mathbf{B}_{\mathcal{W} \mathcal{W}}^{\prime} & -\mathbf{G}_{\mathcal{W} \mathcal{W}} \\
    \hline \mathbf{G}_{\mathcal{W} \mathcal{R}} & \mathbf{G}_{\mathcal{W} \mathcal{W}} & \mathbf{B}_{\mathcal{W} \mathcal{W}}
    \end{array}\right]
\end{align}

In \eqref{DLPF_2}, $\tilde{\mathbf{p}}$ denotes the calculated active power injection vector at sets $\mathcal{R}$ and $\mathcal{W}$, and $\tilde{\mathbf{q}}$ denotes the calculated reactive power injection vector at the PQ buses. They are obtained by the following expression:
\begin{align} \label{DLPF_3}
\begin{bmatrix}
\tilde{\mathbf{p}} \\
\tilde{\mathbf{q}}
\end{bmatrix}
=
\begin{bmatrix}
\mathbf{p}_\mathcal{R} \\
\mathbf{p}_\mathcal{W} \\
\mathbf{q}_\mathcal{W}
\end{bmatrix}
+
\begin{bmatrix}
\mathbf{B}'_{\mathcal{R}\mathcal{S}} & -\mathbf{G}_{\mathcal{R}\mathcal{S}} & -\mathbf{G}_{\mathcal{R}\mathcal{R}} \\
\mathbf{B}'_{\mathcal{W}\mathcal{S}} & -\mathbf{G}_{\mathcal{W}\mathcal{S}} & -\mathbf{G}_{\mathcal{W}\mathcal{R}} \\
\mathbf{G}_{\mathcal{W}\mathcal{S}} & \mathbf{B}_{\mathcal{W}\mathcal{S}} & \mathbf{B}_{\mathcal{W}\mathcal{R}}
\end{bmatrix}
\begin{bmatrix}
\boldsymbol{\theta}_\mathcal{S} \\
\mathbf{v}_\mathcal{S} \\
\mathbf{v}_\mathcal{R}
\end{bmatrix} 
\end{align}
where \textcolor{blue}{$\boldsymbol{\theta}_{\mathcal{S}}$ and $\mathbf{v}_{\mathcal{S}}$ denote the known phase angles and voltage magnitudes in bus set $\mathcal{S}$;} the corresponding entries in $\mathbf{v}_{\mathcal{R}}$ for all the IBR buses are replaced with the collective STAN predictions $\hat{\mathbf{x}}^t$. Therefore, \eqref{DLPF_3} provides the calculated vectors $\tilde{\mathbf{p}}$ and $\tilde{\mathbf{q}}$ corresponding to the predicted voltage-controlled bus magnitudes.

Using the calculated vectors $\tilde{\mathbf{p}}$ and $\tilde{\mathbf{q}}$ by \eqref{DLPF_3}, the phase angle and voltage magnitude $\tilde{\boldsymbol{\theta}}$ and $\tilde{\mathbf{v}}$ are obtained from the following decoupled linear expressions:
\begin{align} \label{DLPF_4}
\begin{bmatrix}
\tilde{\boldsymbol{\theta}} \\
\tilde{\mathbf{v}}
\end{bmatrix}
=
\begin{bmatrix}
\tilde{\mathbf{H}}^{-1}\tilde{\mathbf{p}}-\tilde{\mathbf{H}}^{-1}\mathbf{N} \mathbf{L}^{-1} \tilde{\mathbf{q}} \\
\tilde{\mathbf{L}}^{-1}\tilde{\mathbf{q}}-\tilde{\mathbf{L}}^{-1}\mathbf{M} \mathbf{H}^{-1} \tilde{\mathbf{p}}
\end{bmatrix}
\end{align}
where $\tilde{\mathbf{H}}  =\mathbf{H}-\mathbf{N} \mathbf{L}^{-1} \mathbf{M}$, and $\tilde{\mathbf{L}}  =\mathbf{L}-\mathbf{M} \mathbf{H}^{-1} \mathbf{N} $. 

Equation \eqref{DLPF_4} \textcolor{blue}{provides an analytical voltage solution: $\tilde{\boldsymbol{\theta}}$ and $\tilde{\mathbf{v}}$  are obtained from closed-form linearized expressions. This solution differs from the well-known fast decoupled power flow, which reduces the problem size by ignoring the off-diagonal blocks of the Jacobian matrix and requires iterative runs to obtain the final voltage solution.}

The computed phase angle $\tilde{\boldsymbol{\theta}}$ and voltage magnitude $\tilde{\mathbf{v}}$ are then concatenated as the real-valued network voltage solution vector $\hat{\mathbf{z}}^t=[\tilde{\boldsymbol{\theta}},\tilde{\mathbf{v}}]$ for each timestep. A subset of $\hat{\mathbf{z}}^t$ is utilized by each STAN predictor as additional network information to enhance the predictions by capturing the spatial correlations of trajectory dynamics between neighboring buses. The two sets of voltage solutions, $\hat{\mathbf{x}}^t$ from the STAN-based predictor and $\hat{\mathbf{z}}^t$ from the network operation, are then used in \eqref{MSE_Total} to compute the hybrid physics-informed loss. This facilitates efficient NN gradient computation and backpropagation, accelerating and stabilizing the gradient-based training of the proposed method.

\textcolor{blue}{\textit{Mitigation of Linearization Errors in the Proposed Method.}
The original DL-ACPF formulation assumes near-nominal voltage magnitudes and moderate voltage angle differences. While these assumptions can introduce approximation errors when used alone, the proposed method uses the DL-ACPF solution as a physics-informed regularization term. The proposed hybrid loss function in \eqref{MSE_Total} combines this regularization term with the data-driven loss during end-to-end training, which mitigates the effect of the linearization errors. This is validated by the case studies to some extent, which demonstrate improved system-wide voltage trajectory prediction and physical consistency.}



\vspace{-7pt}
\subsection{Network Interdependency-Informed STAN Gradient Update} \label{Grad}

The proposed STAN for each IBR is trained by calculating both data-driven and physics-informed gradients in the hybrid loss function \eqref{MSE_Total} to update the neural network parameters. 

To inform the network interdependency of the STAN training, the derivatives of the DL-ACPF model are calculated and propagated efficiently as gradients.
In \eqref{DLPF_3}, the STAN predictions directly influence the power injection vectors $\tilde{\mathbf{p}}$ and $\tilde{\mathbf{q}}$, which subsequently influence $\tilde{\boldsymbol{\theta}}$ and $\tilde{\mathbf{v}}$ in \eqref{DLPF_4}. The chain rule is used to bridge the influence of the STAN predictions to the physics-informed loss function by:
\begin{align} \label{BackProp_1}
        \frac{\partial \mathcal{L}_2}{\partial \Phi}&=\frac{\partial \mathcal{L}_2}{\partial \hat{\mathbf{z}}}  \cdot \biggl( \frac{\partial \hat{\mathbf{z}}}{\partial \tilde{\mathbf{p}}} \cdot \frac{\partial \tilde{\mathbf{p}}}{\partial \mathbf{v}_{\mathcal{R}}}  +  \frac{\partial \hat{\mathbf{z}}}{\partial  \tilde{\mathbf{q}}} \cdot \frac{\partial  \tilde{\mathbf{q}}}{\partial \mathbf{v}_{\mathcal{R}}} \biggl) \cdot \frac{\partial \mathbf{v}_{\mathcal{R}}}{\partial \hat{\mathbf{x}}} \cdot\frac{\partial \hat{\mathbf{x}}}{\partial\Phi}
\end{align}
where the Jacobians ${\partial\hat{\mathbf{z}}}/{\partial\tilde{\mathbf{p}}}$  and ${\partial\hat{\mathbf{z}}}/{\partial\tilde{\mathbf{q}}}$ are constant matrices derived from $\tilde{\mathbf{H}}^{-1}$, $\tilde{\mathbf{L}}^{-1}$, $\mathbf{N}$, and $\mathbf{M}$. Specifically, the individual components that comprise the Jacobian for voltage magnitudes and phase angles are expressed as:
\begin{align} \label{Jacobian}
    &\frac{\partial\tilde{\boldsymbol{\theta}}}{\partial\tilde{\mathbf{p}}}=  \tilde{\mathbf{H}}^{-1} , \;\;
    \frac{\partial\tilde{\boldsymbol{\theta}}}{\partial\tilde{\mathbf{q}}} =   -\tilde{\mathbf{H}}^{-1}\mathbf{N}\mathbf{L}^{-1}, \\
    &\frac{\partial\tilde{\mathbf{v}}}{\partial\tilde{\mathbf{p}}} = -\tilde{\mathbf{L}}^{-1}\mathbf{M}{\mathbf{H}}^{-1},\;\;
    \frac{\partial\tilde{\mathbf{v}}}{\partial\tilde{\mathbf{q}}} = \tilde{\mathbf{L}}^{-1}.
\end{align}

Meanwhile, the Jacobians ${\partial\tilde{\mathbf{p}}}/{\partial{\mathbf{v}_{\mathcal{R}}}}$ and ${\partial\tilde{\mathbf{q}}}/{\partial{\mathbf{v}_{\mathcal{R}}}}$ are obtained by applying the matrix operations to \eqref{DLPF_3}:
\begin{align} \label{Jacobian 2}
    \frac{\partial\tilde{\mathbf{p}}}{\partial\mathbf{v}_{\mathcal{R}}} &=  \begin{bmatrix}
        -\mathbf{G}_{\mathcal{RR}} \\ -\mathbf{G}_{\mathcal{WR}}
    \end{bmatrix} \\
    \frac{\partial\tilde{\mathbf{q}}}{\partial\mathbf{v}_{\mathcal{R}}} &= \mathbf{B}_{\mathcal{WR}}
\end{align}

The remaining gradients of \eqref{BackProp_1} can be obtained by computing ${\partial{\mathcal{L}_{2}}}/{\partial{\hat{\mathbf{z}}}}$ from \eqref{PF_MSE}, ${\partial \mathbf{v}_{\mathcal{R}}}/{\partial \hat{\mathbf{x}}}$ (a column vector with non-zero entries only at the IBR bus indices), and  ${\partial{\hat{\mathbf{x}}}}/{\partial{\Phi}}$ from \eqref{FC_layer}.

\textcolor{blue}{The learnable parameters of all STAN predictors are denoted as $\Phi=\{\Phi_1,\Phi_2,\dots,\Phi_{N_{\mathrm{IBR}}}\}$. At training epoch $k$, the joint update is performed by stochastic gradient descent on the hybrid loss function \eqref{MSE_Total}:
\begin{align} \label{SGD_Update}
    \Phi^{(k+1)} = \Phi^{(k)} - \alpha\nabla_{\Phi} \left(\frac{1}{N_{\text{IBR}}} \sum_{j=1}^{N_{\text{IBR}}} \mathcal{L}_{1,j} + \lambda \mathcal{L}_2 \right)
\end{align}}
\textcolor{blue}{where $\alpha$ denotes the learning rate; $\nabla_{\Phi}\mathcal{L}_{2}$ denotes the physics-informed gradient and is obtained by \eqref{BackProp_1}; $\nabla_{\Phi}\mathcal{L}_{1,j}$ denotes the data-driven gradient for the IBR $j$ and is obtained by 
\begin{align} \label{eq: DD_BackProp}
    \frac{\partial \mathcal{L}_{1,j}}{\partial \Phi_j} 
    = 
    \frac{\partial \mathcal{L}_{1,j}}{\partial \hat{\mathbf{x}}_j}
    \frac{\partial \hat{\mathbf{x}}_j}{\partial \Phi_j}
\end{align}
where ${\partial \mathcal{L}_{1,j}}/\partial \hat{\mathbf{x}}_j$ is obtained from \eqref{MSE}, and  ${\partial{\hat{\mathbf{x}}_j}}/{\partial{\Phi}_j}$ follows a standard neural network backpropagation through the proposed STAN architecture.}

\textcolor{blue}{Hence, the parameters of each STAN predictor are updated through its own data-driven gradient from $\mathcal{L}_{1,j}$ and a shared physics-informed gradient from the system-wide loss $\mathcal{L}_2$.} By incorporating physics-driven linearized ACPF, the proposed framework achieves computational efficiency while ensuring system-wide physical consistency in trajectory prediction.

\begin{algorithm}[!htbp]
\caption{Offline Training of Proposed Algorithm}
\label{alg:training_procedure}
\begin{flushleft}
\textbf{Input:} Initial IBR state trajectories $\mathbf{x}^{1:T_P}=\{\mathbf{x}^{1:T_P}_1, \mathbf{x}^{1:T_P}_2, \dots, \mathbf{x}^{1:T_P}_{N_{\text{IBR}}}\}$; learning rate $\alpha$; regularization coefficient $\lambda$\\
\textbf{Output:} STAN parameters $\Phi = \{\Phi_1$, $\Phi_2$, \dots, $\Phi_{N_{\text{IBR}}}$\} 
\end{flushleft}
\begin{algorithmic}[1]
\STATE Initialize all STAN parameters $\Phi$ randomly
\FOR{Each training epoch}
    \FOR{$t = T_P+1$ to $T_{\text{end}}$}
        \FOR{$j=1$ to $N_{\text{IBR}}$}
            \STATE Predict next $T_R$ steps of voltages by \eqref{LSTM_hidden}-\eqref{FC_layer}: \\
            \quad $\hat{\mathbf{x}}^{t:t+T_R-1}_j = \mathbf{f}^{(j)}_{\text{STAN}}(\mathbf{x}_j^{t-T_P:t-1}, \mathbf{y}^{t-T_P:t-1}_{i\in \mathcal{N}_j}; {\Phi}_j)$
        \ENDFOR
        \STATE Compute  $\mathbf{z}^{t:t+T_R-1}$ and $\hat{\mathbf{z}}^{t:t+T_R-1}$ by \eqref{DLPF_4} 
        \STATE Compute the data-driven loss $\{\mathcal{L}_{1,j}\}_{j=1}^{N_{\text{IBR}}}$ by \eqref{MSE} and physics-informed $\mathcal{L}_2$ by \eqref{PF_MSE}
        \STATE Update $\Phi=\{\Phi_j\}_{j=1}^{N_{\text{IBR}}}$ using \eqref{SGD_Update} by combining the DL-ACPF gradients by \eqref{BackProp_1} and data-driven gradients by \eqref{eq: DD_BackProp}.
    \ENDFOR
    \IF{convergence or maximum iterations reached}
    \STATE \textbf{break}
    \ENDIF
\ENDFOR
\end{algorithmic}
\end{algorithm}

\vspace{-5pt}
\subsection{Multi-Timestep and System-Wide Trajectory Prediction} \label{sec: System-Wide}

As shown in Fig. \ref{fig:Architecture}, the proposed STAN predictors serve as a data-driven surrogate model on each IBR bus, generating voltage predictions at each timestep. Each STAN predictor performs multi-step prediction by \eqref{Bbox_multstep} recursively for the trajectory length up to $T_{\text{end}}$. The pseudocode for the training procedure of the proposed framework is shown in Algorithm \ref{alg:training_procedure}.

\textcolor{blue}{The proposed framework is trained in a joint end-to-end manner across all $N_{\text{IBR}}$ STAN predictors, and the predicted IBR trajectories jointly produce a physically consistent system-wide voltage trajectory.
During offline training, the parameters of all $N_{\text{IBR}}$ STAN predictors are jointly updated by \eqref{SGD_Update}. }

\begin{figure}
    \centering
    \includegraphics[width=\linewidth]{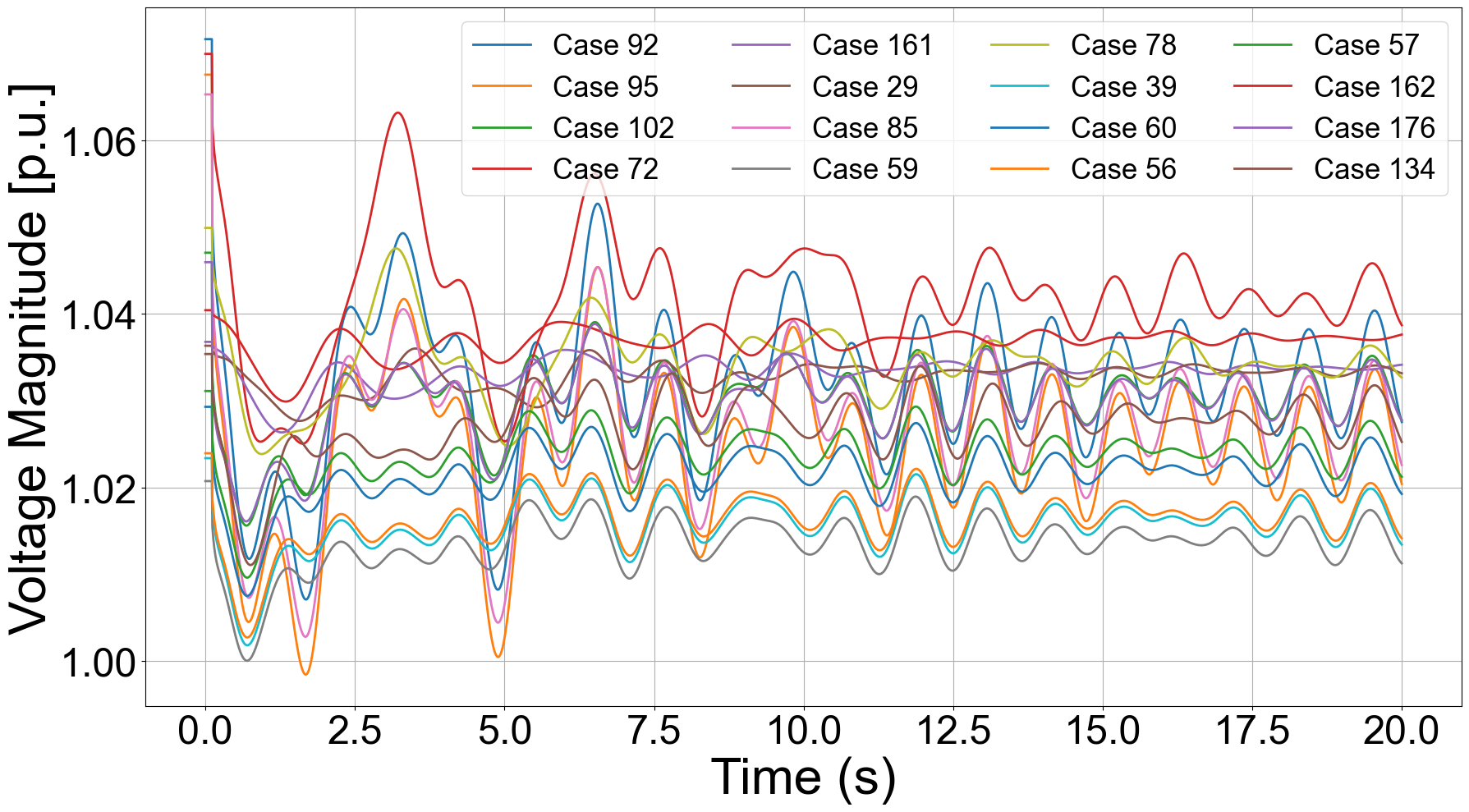}
    \caption{16 randomly selected \textcolor{blue}{ground-truth} voltage magnitude trajectories of an IBR bus under fault in the WECC 179-bus system.}
    \label{fig:Trajectory_Comp}
\end{figure}

\section{Case Studies}

The proposed algorithm is tested on the IEEE 14-bus and WECC 179-bus systems \cite{Maslennikov7741772}, which are modified by adding droop-controlled IBRs in ANDES \cite{HantaoCui9169830}. Comprehensive control schemes, such as turbine governors and excitation systems, are adopted to model SGs and IBRs. To simulate various contingency types, generator faults and load shedding are randomly set in different locations and occur at $0.1$ s. The base loads are randomly varied between $80\%$ and $120\%$ to simulate different system loading conditions. The PMU measurement noise is considered, and the maximum errors are set as 1\% \cite{PMUerror8815851}.
The initial input to the proposed algorithm consists of the initial post-fault voltage trajectories of all IBR buses over $2.0$ s, corresponding to $120$ timesteps. \textcolor{blue}{The remaining $18.0$ s trajectory is predicted recursively from this initial $2.0$ s input window, without updating the model inputs using ground-truth measurements during the prediction horizon.} The total simulation time is $20$ s with a sampling rate of $60$ Hz. 
Fig. \ref{fig:Trajectory_Comp} exemplifies \textcolor{blue}{simulated ground-truth voltage} magnitude trajectories generated in the WECC 179-bus system for the offline training.

The performance of the proposed algorithm is compared to three benchmark methods: a purely data-driven LSTM network, a recent physics-informed ML approach \cite{IntegratingLiJiaming10066348} (namely, PF-Integrated ML), and \textcolor{blue}{a graph neural network (GNN)-based model}. The PF-Integrated ML method is adapted as an LSTM black box for IBR buses that \textcolor{blue}{requires complete network information to} incorporate power flow solutions as inputs to predict their dynamic trajectories. \textcolor{blue}{The GNN-based method is used to evaluate the performance of using network-structured voltage information to capture spatial dependencies.} The trajectory prediction accuracy is evaluated by using the root mean square errors (RMSEs) and mean absolute errors (MAEs) on the testing dataset. Meanwhile, the maximum absolute error is used to assess robustness to measurement errors.

 A total of $4,000$ trajectories are generated for each fault scenario, which is partitioned randomly into training ($70\%$), validation ($20\%$), and testing ($10\%$) subsets. The proposed STAN-based predictor consists of two LSTM layers with 128 and 64 hidden units, and an attention layer with 64 hidden units. The method is trained over 100 epochs using the Adam optimizer with a learning rate of 0.001. \textcolor{blue}{The regularization coefficient $\lambda$ in \eqref{MSE_Total} is selected using a validation-based coarse-to-fine grid search. 
Table \ref{tab:lambda_sensitivity_both} summarizes the sensitivity analysis of $\lambda$, where $\lambda=0$ corresponds to training without the physics-informed loss term. The selected values of $\lambda$ are $0.07$ for the IEEE 14-bus system and $0.05$ for the 179-bus test system, which yield the lowest validation RMSEs for voltage magnitude and phase angle.}

\textcolor{blue}{To further examine the separate contribution of attention-based temporal feature extraction and physics-informed structure, an ablation study is provided in Section \ref{sec: Ablation}}.
All simulations and training procedures are executed on a desktop computer equipped with an Intel Core i9-10900K 3.70-GHz processor and 32 GB of RAM.

\begin{table}[!t]
  \centering
  \caption{\textcolor{blue}{\textsc{Sensitivity Analysis of the Physics-Informed Loss Weight $\lambda$}}}
  \label{tab:lambda_sensitivity_both}
  \setlength{\tabcolsep}{4pt}

  \begin{subtable}[t]{0.47\columnwidth}
  \centering
  \caption{IEEE 14-bus system}
  \label{tab:lambda_sensitivity_ieee14}
  \begin{tabular}{c c c}
  \hline
  $\lambda$ & $|V|$ RMSE & $\theta$ RMSE \\
  \hline
  0    & 1.94e-03 & 2.50e-03 \\
  0.01 & 1.76e-03 & 2.41e-03 \\
  0.05 & 1.18e-03 & 2.09e-03 \\
  \textbf{0.07} & \textbf{1.01e-03} & \textbf{1.95e-03} \\
  0.10 & 1.11e-03 & 2.04e-03 \\
  \hline
  \end{tabular}
  \end{subtable}
  \hfill
  \begin{subtable}[t]{0.47\columnwidth}
  \centering
  \caption{WECC 179-bus system}
  \label{tab:lambda_sensitivity_wecc179}
  \begin{tabular}{c c c}
  \hline
  $\lambda$ & $|V|$ RMSE & $\theta$ RMSE \\
  \hline
  0    & 3.18e-03 & 5.13e-03 \\
  0.01 & 2.72e-03 & 4.56e-03 \\
  \textbf{0.05} & \textbf{2.29e-03} & \textbf{4.09e-03} \\
  0.07 & 2.40e-03 & 4.55e-03 \\
  0.10 & 2.45e-03 & 4.47e-03 \\
  \hline
  \end{tabular}
  \end{subtable}

  \vspace{1mm}
  \end{table}

\begin{figure}
    \centering

    \begin{subfigure}{\linewidth}
        \centering
        \includegraphics[width=0.87\linewidth]{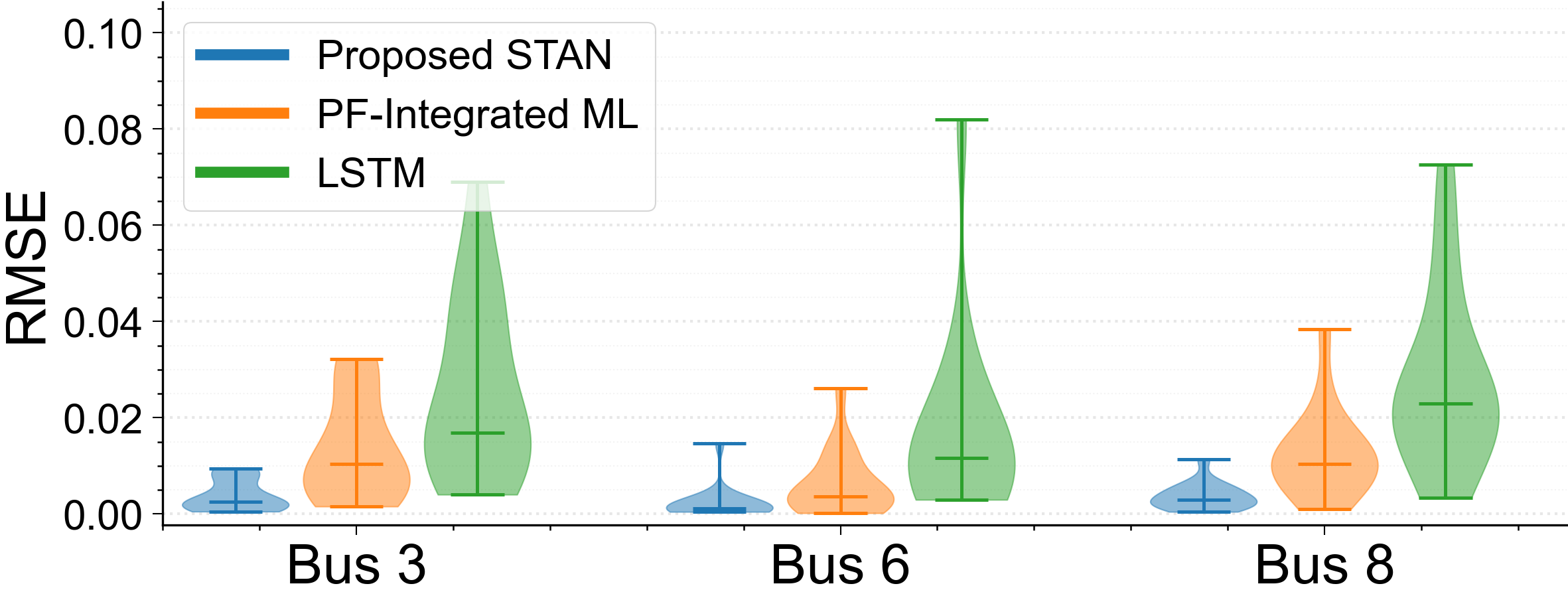}
       \vspace{-5pt}
        \caption{Voltage magnitude}
        \label{fig:IEEE14_dist_comp_vmag}
    \end{subfigure}
    
    \vspace{1em}

    \begin{subfigure}{\linewidth}
        \centering
        \includegraphics[width=0.87\linewidth]{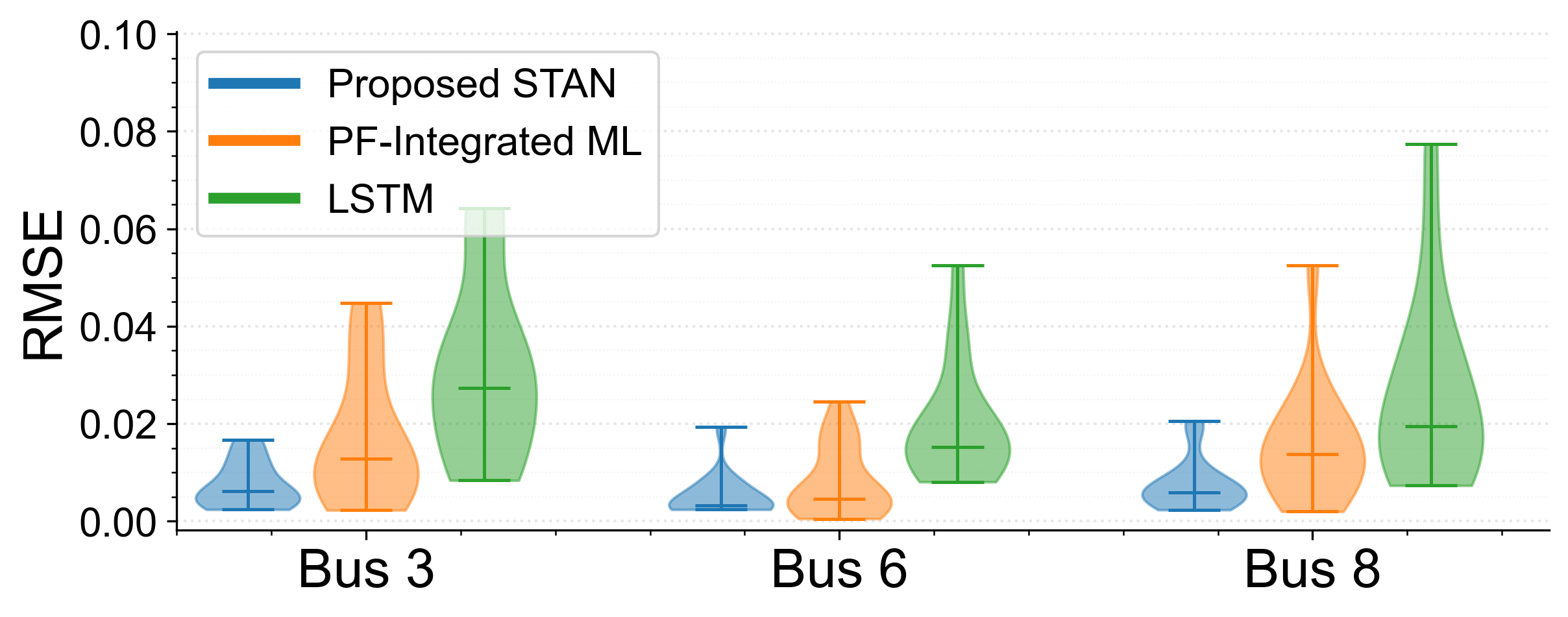}
     \vspace{-5pt}
        \caption{Phase angle}
        \label{fig:IEEE14_dist_comp_ang}
    \end{subfigure}

    \caption{Per-trajectory RMSE distribution comparison of all the IBR buses in the IEEE 14-bus system for 400 test cases.}
    \label{fig:IEEE14_dist_comp_combined}
\end{figure}

\subsection{Case I: IEEE 14-Bus System}

The IEEE 14-bus system is modified by installing IBRs on buses $3$, $6$, and $8$. Fig. \ref{fig:IEEE14_dist_comp_combined} compares the RMSE distributions of per-trajectory voltage magnitude and phase angle on the three IBR buses. The proposed method consistently achieves the lowest median RMSE across all IBR buses, yielding up to an $8$-fold improvement over the LSTM model. For example, the median phase angle RMSE of the proposed method on bus $3$ is $7.48 \times 10^{-3}$, which is nearly three times lower than that of LSTM at $2.04 \times 10^{-2}$. The PF-Integrated ML method improves over the LSTM baseline but still lags behind the proposed method. For instance, the median voltage magnitude RMSE of the proposed method on bus $3$ is $2.86 \times 10^{-3}$, approximately three times lower than that of the PF-Integrated method of $7.92 \times 10^{-3}$. Across all buses, the proposed method also produces the narrowest error distributions. For example, the phase angle RMSE of the proposed method at bus 8 has a range of $2.45 \times 10^{-2}$ with a standard deviation of $5.81 \times 10^{-3}$. Compared with the RMSE range and standard deviation of the PF-Integrated method of $5.24 \times 10^{-2}$ and $9.95 \times 10^{-3}$, respectively, this indicates a $2$-fold enhancement in prediction stability.

\begin{table}[!t]
\centering
\caption{\textsc{Comparative Prediction Performance in the IEEE 14-bus System}}
\label{tab:IEEE14_comp}
\begin{tabular}{ccccc}
\hline
\multirow{2}{*}{\#Method} & \multicolumn{2}{c}{Voltage Magnitude} & \multicolumn{2}{c}{Phase Angle} \\
& RMSE & MAE & RMSE & MAE \\
\hline
\textbf{Proposed} & \textbf{1.08e-03} & \textbf{7.52e-04} & \textbf{2.18e-03} & \textbf{1.48e-03} \\
PF-Integrated ML \cite{IntegratingLiJiaming10066348} & 4.86e-03 & 3.62e-03 & 8.73e-03 & 6.42e-03 \\
Data-driven LSTM & 8.95e-03 & 6.71e-03 & 1.47e-02 & 1.13e-02 \\
\textcolor{blue}{GNN} & \textcolor{blue}{4.21e-03} & \textcolor{blue}{3.08e-03} & \textcolor{blue}{7.52e-03} & \textcolor{blue}{5.51e-03} \\
\hline
\end{tabular}
\end{table}

Table \ref{tab:IEEE14_comp} compares the RMSEs and MAEs of voltage magnitudes and phase angles on the IBR buses in 400 random test cases. The proposed algorithm achieves the lowest errors, with voltage magnitude RMSE and MAE of $1.08 \times10^{-3}$ and $7.52 \times 10^{-4}$, and phase angle RMSE and MAE of $2.18 \times10^{-3}$ and $1.48 \times10^{-3}$, respectively. In contrast, the PF-Integrated ML method produces RMSE and MAE values approximately 5 times higher.
Meanwhile, the LSTM baseline exhibits up to a 12-fold increase in RMSE and MAE values, highlighting the limitations of purely data-driven methods. 
\textcolor{blue}{The GNN-based method improves over the data-driven LSTM. However, compared with the GNN-based method, the proposed method further significantly reduces the RMSEs of voltage magnitude and phase angle.} These results demonstrate that the proposed physics-informed ML method substantially enhances the accuracy of predictions.

\subsection{Case II: WECC 179-Bus System}

\begin{figure}
    \centering
    \begin{subfigure}{\columnwidth}
        \centering
        \includegraphics[width=0.99\linewidth]{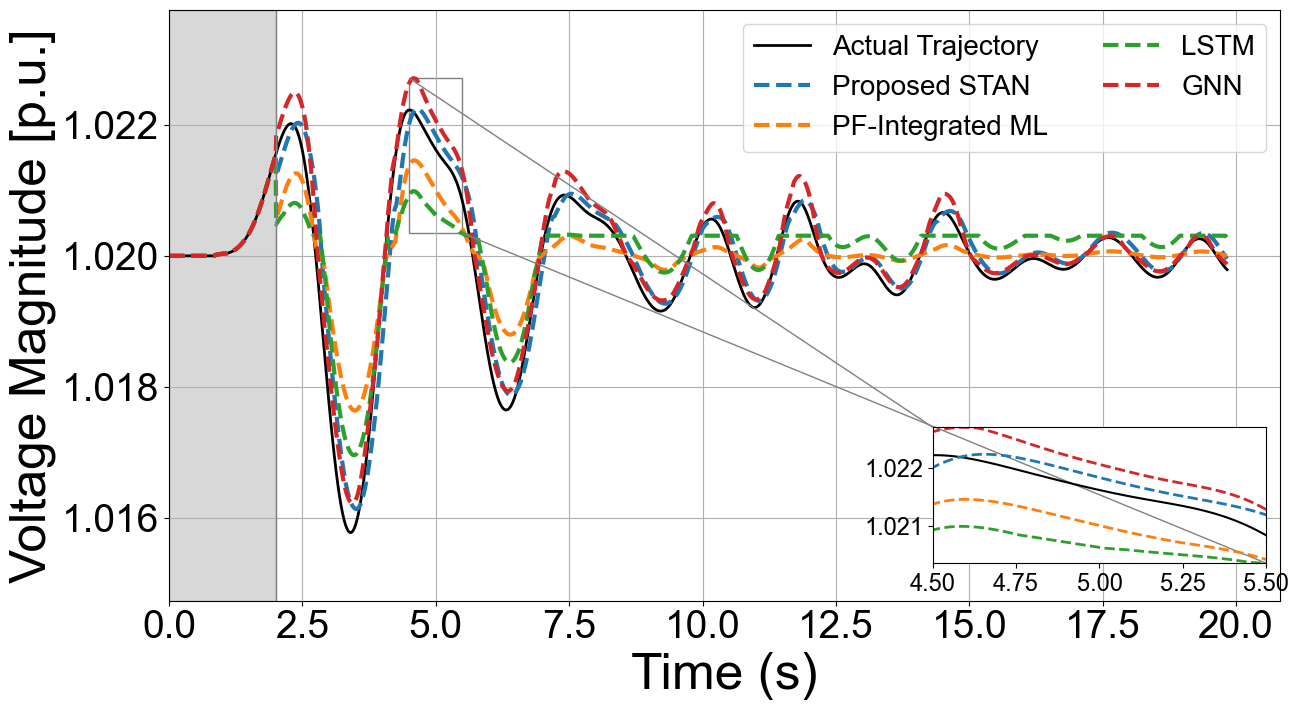}

        \caption{Voltage Magnitude}
        
    \end{subfigure}

    \begin{subfigure}{\columnwidth}
        \centering
        \includegraphics[width=0.99\linewidth]{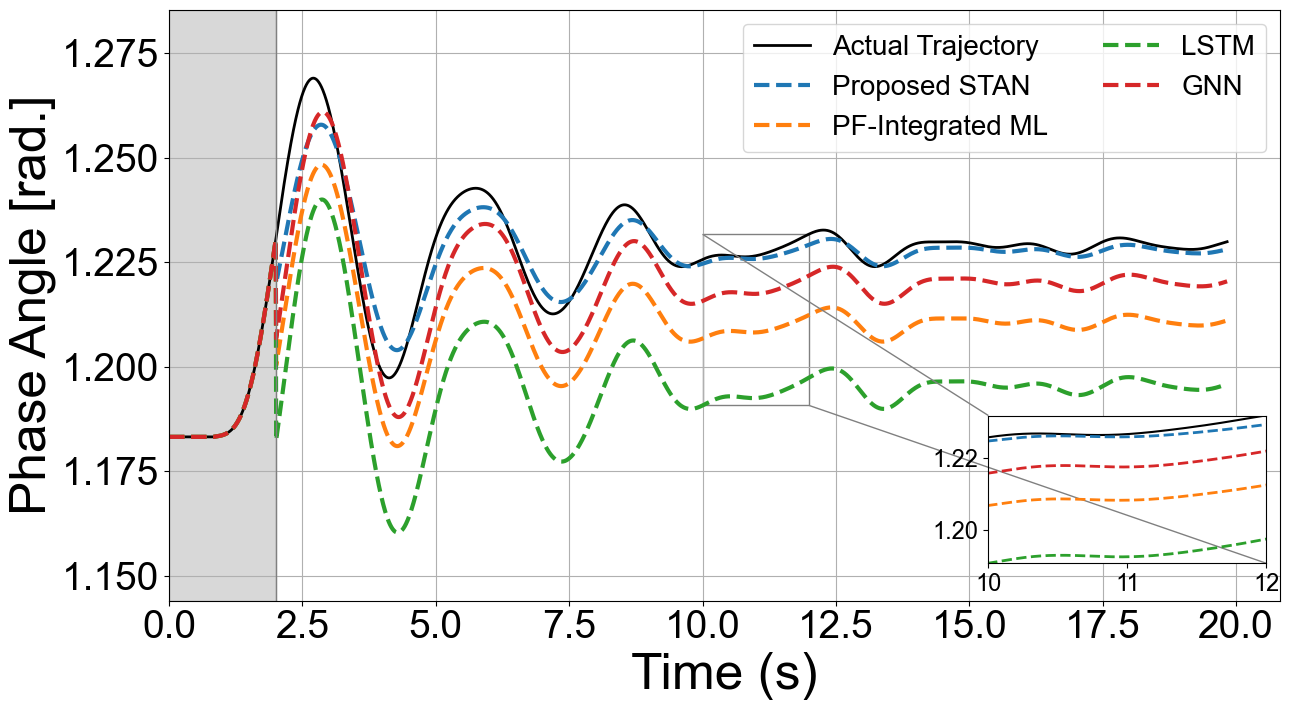}

        \caption{Phase Angle}
    \end{subfigure}
    
    \caption{Trajectory prediction comparison between the proposed method, LSTM, PF-Integrated ML, and \textcolor{blue}{GNN} on IBR 3 in the WECC 179-bus system under a fault.}
    \label{WECC_Traj_Comp}
\end{figure}

To evaluate the performance of the proposed method on a larger-scale IBR-integrated power grid, the original WECC 179-bus system is modified by installing 10 IBRs on buses 8, 10, 12, 17, 29, 34, 39, 46, 64, and 117. 

\subsubsection{IBR Bus Trajectory Prediction}
The predicted dynamic trajectories of IBR 3 (bus 12) for the proposed method and benchmark methods are illustrated in Fig. \ref{WECC_Traj_Comp}. The shaded gray area represents the initial trajectory provided to the model as input. It is shown that the proposed method closely follows the actual voltage magnitude and phase angle trajectories. The LSTM method exhibits noticeable deviations due to its lack of network-aware physical information, especially for the phase angle trajectory. The PF-Integrated ML and GNN-based methods outperform the LSTM baseline, but both still exhibit visible deviations from the actual response in the zoomed-in regions. \textcolor{blue}{The GNN-based method captures the general trend but often overestimates several voltage magnitude peaks and does not match the phase angle trajectory as closely as the proposed method. These trajectories show that the proposed method provides the most accurate IBR-bus trajectory prediction among the compared methods.}

Table \ref{tab:WECC_all} compares the prediction errors of different methods on IBR buses in the 400 test cases, and an unknown fault is randomly set in each of these cases. The proposed method consistently achieves the lowest errors for voltage magnitude and phase angle.
These results highlight the effectiveness of the proposed method in achieving accurate trajectory prediction on a larger-scale IBR-integrated power grid. 

\begin{table*}[!t]
\centering
\caption{\textsc{Comparative Prediction Performance on All the Buses in the WECC 179-bus System}}
\label{tab:WECC_all}
\renewcommand{\arraystretch}{1.1}
\setlength{\tabcolsep}{3pt}
\resizebox{\textwidth}{!}{%
\begin{tabular}{cccccccccc}
\hline
\multirow{2}{*}{Method} &
\multicolumn{3}{c}{Voltage Magnitude (p.u.)} &
\multicolumn{3}{c}{Phase Angle (rad.)} &
\multirow{2}{*}{\shortstack{Unseen Scenario\\RMSE (V/$\theta$)}} &
\multirow{2}{*}{\shortstack{System-wide\\RMSE (V/$\theta$)}} \\ 
\cline{2-7}
 & RMSE/MAE & wErr RMSE & MaxErr &
   RMSE/MAE & wErr RMSE & MaxErr & & \\ 
\hline

\textbf{Proposed Method} & 
\textbf{2.37e-03 / 1.69e-03} & \textbf{2.69e-03} & \textbf{4.51e-03} &
\textbf{4.15e-03 / 2.61e-03} & \textbf{4.68e-03} & \textbf{9.25e-03} &
\textbf{4.86e-03 / 7.48e-03} &
\textbf{2.96e-03 / 5.06e-03} \\

PF-Integrated ML \cite{IntegratingLiJiaming10066348} & 
9.92e-03 / 7.15e-03 & 1.77e-02 & 5.82e-02 &
1.71e-02 / 1.29e-02 & 4.12e-02 & 1.22e-01 &
2.56e-02 / 4.49e-02 &
1.30e-02 / 2.57e-02 \\

Data-driven LSTM & 
1.93e-02 / 1.68e-02 & 4.85e-02 & 2.39e-01 &
4.18e-02 / 3.57e-02 & 6.68e-02 & 4.35e-01 &
9.07e-02 / 1.40e-01 &
4.32e-02 / 7.17e-02 \\

\textcolor{blue}{GNN} &
\textcolor{blue}{7.85e-03 / 5.62e-03} & \textcolor{blue}{1.21e-02} & \textcolor{blue}{3.94e-02} &
\textcolor{blue}{1.38e-02 / 9.84e-03} & \textcolor{blue}{2.86e-02} & \textcolor{blue}{9.76e-02} &
\textcolor{blue}{1.86e-02 / 3.21e-02} &
\textcolor{blue}{9.72e-03 / 1.91e-02} \\

\hline
\end{tabular}%
}
\end{table*}

{\color{blue}
In addition to metrics on voltage magnitude and phase angle, 
the active and reactive power injection errors across all IBR buses are compared in Table \ref{tab:WECC_PQ_Performance}. The proposed method achieves the lowest active power RMSE and MAE of $1.43\times 10^{-2}$ and $3.90\times 10^{-3}$, respectively. Compared with the PF-Integrated ML, GNN, and data-driven LSTM methods, the proposed method reduces the active power RMSE by approximately 27\%, 24\%, and 64\%, respectively.
The proposed method achieves similar superiority in reactive power injection and achieves the lowest RMSE and MAE.
These results demonstrate that the proposed framework provides a more comprehensive prediction of IBR dynamic responses by accurately representing both voltage trajectories and IBR power-injection behavior.

}

\begin{table}[!t]
\centering
\caption{\textsc{\textcolor{blue}{IBR Power Injection Evaluation in WECC 179-Bus System}}}
\label{tab:WECC_PQ_Performance}
\begin{tabular}{ccccc}
\hline
\multirow{2}{*}{\#Method} & \multicolumn{2}{c}{Active Power} & \multicolumn{2}{c}{Reactive Power} \\
& RMSE & MAE & RMSE & MAE \\
\hline
\textbf{Proposed STAN} & \textbf{1.43e-02} & \textbf{3.90e-03} & \textbf{2.21e-02} & \textbf{6.54e-03} \\
PF-Integrated ML \cite{IntegratingLiJiaming10066348} & 1.95e-02 & 4.74e-03 & 2.72e-02 & 6.84e-03 \\
Data-driven LSTM & 4.01e-02 & 6.43e-03 & 2.79e-02 & 7.38e-03 \\
\textcolor{blue}{GNN} & \textcolor{blue}{1.89e-02} & \textcolor{blue}{4.40e-03} & \textcolor{blue}{2.61e-02} & \textcolor{blue}{6.77e-03} \\
\hline
\end{tabular}
\end{table}

\subsubsection{Adaptability to Unseen Scenarios} To evaluate the adaptability of the proposed method and benchmark methods to unseen scenarios, all methods are tested on scenarios that were excluded from the original training dataset: 1) enlarged load varying range from $70\%$ to $130\%$, compared to that in offline training, 2) generator $5$ is suddenly tripped, which was deliberately excluded in the generation of the training dataset. A total of $100$ new cases are generated as unseen scenarios.

\begin{figure}
    \centering
    \includegraphics[width=\linewidth]{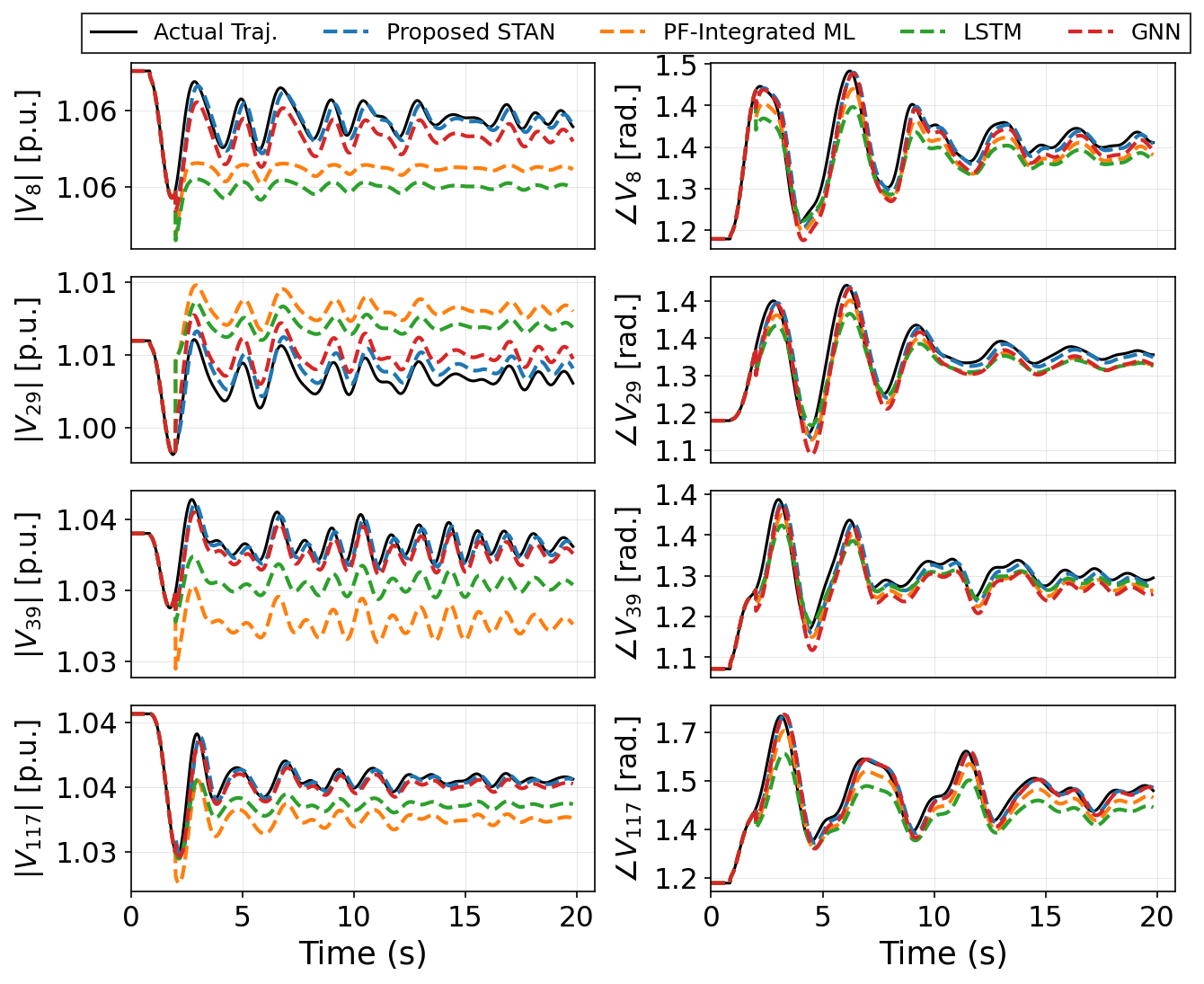}
    \caption{Trajectory prediction comparison between the proposed method, LSTM, PF-Integrated ML, and \textcolor{blue}{GNN} on multiple IBR buses in the WECC 179-bus system under an unseen scenario.}
    \label{fig: unseen_scenario_traj_comp}
\end{figure}

Fig. \ref{fig: unseen_scenario_traj_comp} compares the trajectory prediction results on buses $8$, $29$, $39$, and $117$ (IBRs $1$, $5$, $7$, and $10$) under an unseen scenario. The proposed method closely tracks the actual voltage magnitude and phase angle trajectories across these IBR buses, while the benchmark methods exhibit larger deviations in different portions of the response. The LSTM baseline shows the most pronounced offsets, particularly in the voltage magnitude trajectories, indicating limited adaptability to the unseen operating condition. \textcolor{blue}{The PF-Integrated ML and GNN-based methods improve over the LSTM baseline and capture the overall dynamic trends, but they still show visible amplitude and steady-state deviations from the actual trajectories. In comparison, the proposed method maintains closer agreement with the actual responses across both voltage magnitude and phase angle, visually confirming its stronger adaptability to unseen scenarios.}

Table \ref{tab:WECC_all} also lists the voltage magnitude and phase angle RMSEs of the proposed method and benchmark methods under unseen scenarios. The proposed method demonstrates the strongest adaptability to unseen scenarios, achieving the lowest RMSE and exhibiting the lowest performance degradation compared to evaluation on scenarios included in the training dataset. For instance, the proposed method experiences a $105\%$ performance degradation in voltage magnitude, while the PF-Integrated and LSTM methods experience $160\%$ and $370\%$ degradations, respectively. 
\textcolor{blue}{The GNN-based method achieves unseen-scenario voltage magnitude and phase angle RMSEs of $1.86\times10^{-2}$ and $3.21\times10^{-2}$, respectively. 
} \textcolor{blue}{These observations confirm that the proposed network interdependency-informed framework improves adaptability under unseen operating conditions. 
}

\subsubsection{Overall Trajectory Prediction across Grid}
The comparative prediction performance on all the buses is also summarized in Table \ref{tab:WECC_all}. The proposed method consistently outperforms the PF-Integrated and LSTM benchmarks, achieving significant accuracy enhancement in voltage magnitude and phase angle RMSE, respectively. Specifically, the proposed method achieves voltage magnitude and phase angle RMSEs of $2.96\times10^{-3}$ and $5.06\times10^{-3}$, compared with $1.30\times10^{-2}$ and $2.57\times10^{-2}$ for the PF-Integrated ML method, and $4.32\times10^{-2}$ and $7.17\times10^{-2}$ for the data-driven LSTM. \textcolor{blue}{
The proposed method further reduces these errors to $2.96\times10^{-3}$ and $5.06\times10^{-3}$, corresponding to about $70\%$ and $74\%$ reductions relative to the GNN-based method.} These results verify the ability of the proposed method to predict accurate system-wide voltage trajectories in the IBR-integrated grid.

\subsection{Robustness against Measurement Errors}

This section focuses on evaluating the robustness of the proposed method and benchmark methods against measurement errors. The voltage trajectory prediction errors under PMU measurement noise (denoted as wErr RMSE) in the WECC 179-bus system are presented in Table \ref{tab:WECC_all}. The proposed method exhibits the highest robustness against PMU measurement noise, whereas the benchmark models experience significant degradation in prediction accuracy. Specifically, the proposed method demonstrates only a $13\%$ increase in the RMSE of voltage magnitudes and phase angles under measurement errors. Meanwhile, the PF-Integrated and LSTM models exhibit significantly higher susceptibility to measurement errors. \textcolor{blue}{The GNN-based method is more robust than the data-driven LSTM but remains less robust than the proposed method, with wErr RMSEs of $1.21\times10^{-2}$ and $2.86\times10^{-2}$ for voltage magnitude and phase angle, respectively. The proposed method reduces these errors by approximately $78\%$ and $84\%$, respectively, relative to the GNN-based method.}

\begin{table}[!t]
\centering
\caption{\textcolor{blue}{\textsc{Evaluation and Comparison of Physical Consistency of
Different Methods}}}
\label{tab:physical_consistency}
\setlength{\tabcolsep}{3pt}
\renewcommand{\arraystretch}{1.08}

\begin{subtable}[t]{\columnwidth}
\centering
\caption{IEEE 14-bus System}
\label{tab:physical_consistency_ieee14}
\begin{tabular}{cccc}
\hline
Method & Nodal $P$ & Nodal $Q$ & Branch $|I|$ \\
\hline
\textbf{Proposed} & \textbf{1.07e-01} & \textbf{2.45e-01} & \textbf{7.74e-02} \\
PF-Integrated ML \cite{IntegratingLiJiaming10066348} & 1.43e-01 & 2.75e-01 & 8.59e-02 \\
Data-driven LSTM & 1.54e-01 & 2.65e-01 & 9.10e-02 \\
\textcolor{blue}{GNN} & \textcolor{blue}{1.35e-01}& \textcolor{blue}{2.74e-01}& \textcolor{blue}{8.51e-02}\\
\hline
\end{tabular}
\end{subtable}

\vspace{2mm}

\begin{subtable}[t]{\columnwidth}
\centering
\caption{WECC 179-bus System}
\label{tab:physical_consistency_wecc179}
\begin{tabular}{cccc}
\hline
Method & Nodal $P$ & Nodal $Q$ & Branch $|I|$ \\
\hline
\textbf{Proposed} & \textbf{2.11e-02} & \textbf{3.13e-02} & \textbf{1.69e-02} \\
PF-Integrated ML \cite{IntegratingLiJiaming10066348} & 4.36e-02& 4.54e-02 & 2.24e-02 \\
Data-driven LSTM & 5.82e-02& 5.83e-02 & 3.08e-02 \\
\textcolor{blue}{GNN} & \textcolor{blue}{3.74e-02}& \textcolor{blue}{4.38e-02}& \textcolor{blue}{1.99e-02}\\
\hline
\end{tabular}
\end{subtable}
\vspace{3mm}
\footnotesize{All entries are RMSE values.}
\end{table}

Furthermore, the maximum absolute error (denoted as MaxErr) is used to provide a more comprehensive assessment of the models' robustness to measurement errors in Table \ref{tab:WECC_all}. This metric captures the worst-case trajectory deviation across the test dataset, highlighting conditions where measurement errors most severely impact prediction accuracy. The proposed method achieves maximum absolute errors of $4.51\times10^{-3}$ for the voltage magnitude and $9.25\times10^{-3}$ for the phase angle. 
By comparison, the PF-Integrated and LSTM methods exhibit $3$- and $6$-fold performance degradations. \textcolor{blue}{The GNN-based method has maximum absolute errors of $3.94\times10^{-2}$ and $9.76\times10^{-2}$ for voltage magnitude and phase angle, respectively. While the GNN-based method outperforms the LSTM baseline, the proposed method still provides the lowest worst-case trajectory deviations under noisy PMU measurements.}


\textcolor{blue}{Beyond voltage trajectory errors, the physical consistency of the predicted voltages with respect to network equations is evaluated under noisy PMU measurements. Table \ref{tab:physical_consistency} reports the RMSEs of nodal active/reactive power and branch-current magnitudes under 1\% PMU measurement noise for the IEEE 14-bus and WECC 179-bus systems, respectively. These quantities are computed from the predicted voltage through the network equations. The proposed method achieves the lowest errors across these metrics in both systems, showing that it preserves network-level physical consistency more accurately than other methods under noisy measurements. These results demonstrate the comprehensive robustness of the proposed method in noisy PMU environments. The ablation study in Section \ref{sec: Ablation} further indicate that this robustness benefits from the proposed network-informed design.}

{\color{blue}
\subsection{Ablation Study} \label{sec: Ablation}
To evaluate the contribution of physics-informed components in the proposed framework, an ablation study is conducted on the WECC 179-bus system. The full proposed method is compared with two variants: 1) purely data-driven STAN: it removes the network physics-informed loss and neighboring-bus network information to evaluate the value of network interdependency, and 2) physics-informed LSTM: only the attention mechanism is removed in the proposed method. Hence, the comparative analysis of these two variants can clearly highlight the contributions of the proposed physics-informed neural network structure. 

Table~\ref{tab:ablation} compares the voltage magnitude and phase angle prediction errors of the full proposed method and its two variants under 1\% PMU measurement noise. The proposed method achieves the lowest errors across all metrics. Compared with the purely data-driven STAN, the proposed method reduces the voltage magnitude RMSE from $9.73\times10^{-3}$ to $2.72\times10^{-3}$ and the phase angle RMSE from $9.82\times10^{-2}$ to $2.09\times10^{-2}$, corresponding to reductions of nearly $72\%$ and $79\%$, respectively. This indicates that incorporating neighboring-bus information and the network-dependent physics-informed loss is critical for improving prediction accuracy under noisy measurements. The comparison with the physics-informed LSTM further shows the benefit of the attention mechanism. 
These results suggest that the performance enhancement is primarily attributable to the network-aware, physics-informed design.

\begin{table}[!t]
\centering
\caption{\textsc{\textcolor{blue}{Comparison in Voltage Prediction Accuracy with Variants}}}
\label{tab:ablation}
\begin{tabular}{ccccc}
\hline
\multirow{2}{*}{Ablation study} & \multicolumn{2}{c}{Voltage Magnitude} & \multicolumn{2}{c}{Phase Angle} \\
& RMSE & MAE & RMSE & MAE \\
\hline
\textbf{Proposed} & \textbf{2.72e-03} & \textbf{1.82e-03} & \textbf{2.09e-02} & \textbf{1.54e-02} \\
Purely data-driven STAN & 9.73e-03 & 6.57e-03 & 9.82e-02 & 7.68e-02 \\
Physics-informed LSTM & 3.09e-03 & 2.05e-03 & 2.57e-02 & 1.60e-02 \\
\hline
\end{tabular}
\end{table}

\begin{figure}
    \centering
    \begin{subfigure}{\columnwidth}
        \centering
        \includegraphics[width=0.9\linewidth]{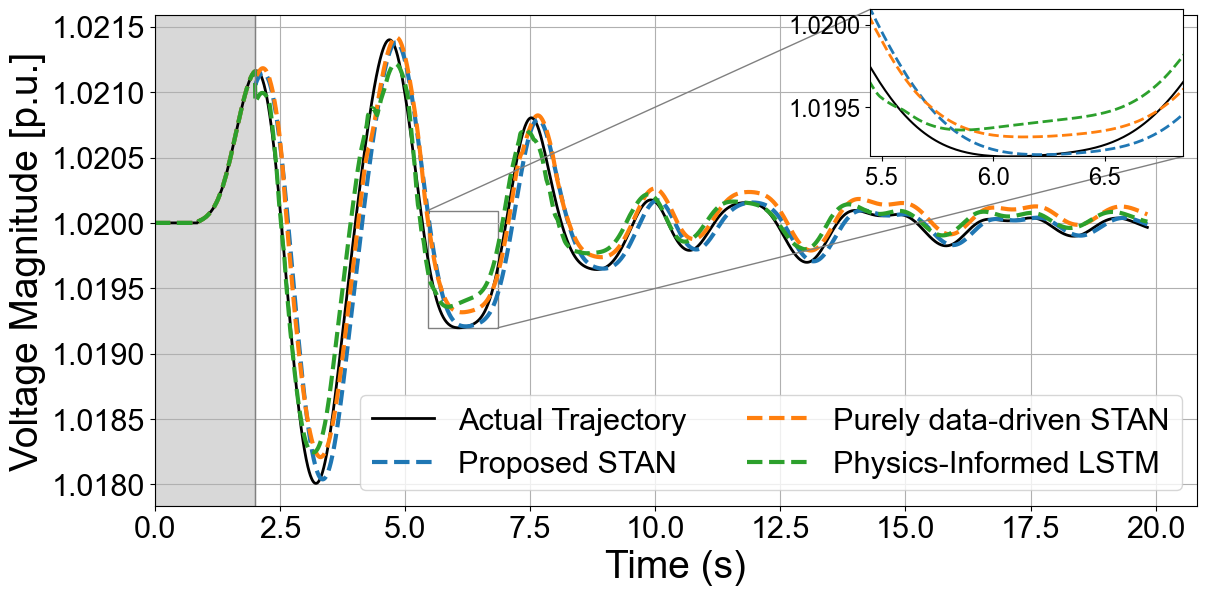}
        \caption{Voltage Magnitude}
        
    \end{subfigure}

    \begin{subfigure}{\columnwidth}
        \centering
        \includegraphics[width=0.9\linewidth]{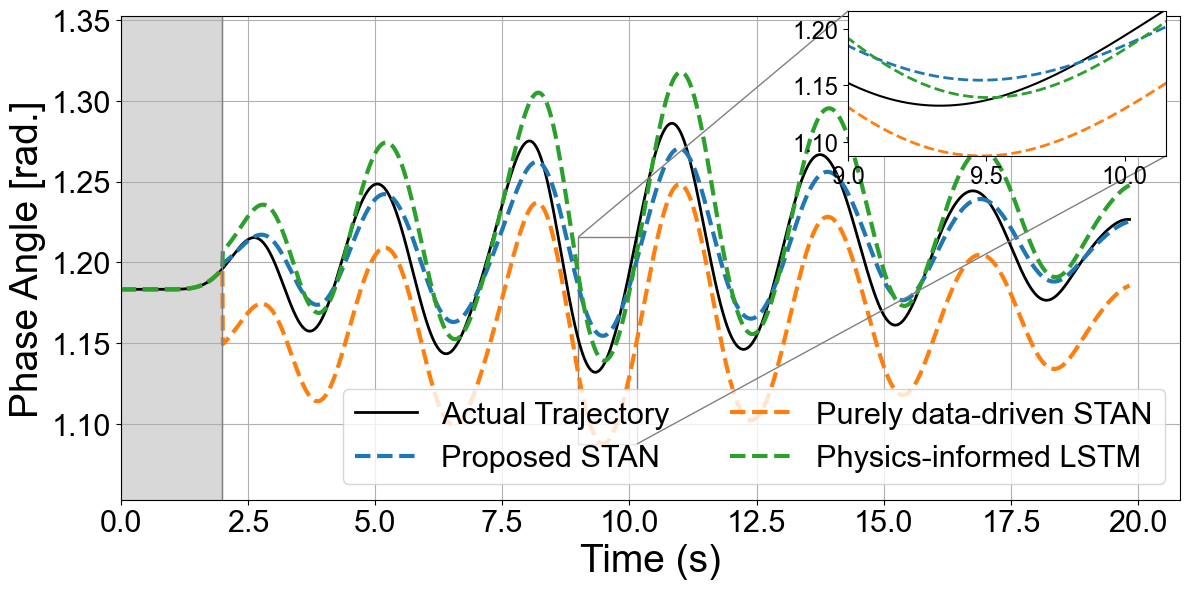}
        \vspace{-8pt}
        \caption{Phase Angle}
    \end{subfigure}
    
    \caption{\textcolor{blue}{Trajectory prediction comparison of the proposed method and its variants on IBR bus 12 in the WECC 179-bus system}}
    \label{Fig: Ablation_Traj_Comp}
\end{figure}


Fig. \ref{Fig: Ablation_Traj_Comp} illustrates the detailed results of voltage magnitude and phase angle trajectories on an IBR bus. The purely data-driven STAN and physics-informed LSTM variants exhibit larger deviations from the actual response than the full proposed method. These trajectory-level results visually support the quantitative comparison in Table \ref{tab:ablation}, demonstrating that the proposed method provides the most accurate dynamic trajectory prediction.
}

\subsection{Computational Efficiency}

To evaluate the computational efficiency for online deployment, the online inference time and memory usage of the proposed method are compared with those of the benchmark methods on the WECC 179-bus system. Table \ref{tab:comput_eff} lists the average results over $100$ test instances. While the LSTM model achieves the lowest inference time of $0.63$ s and memory usage of $1020$ MB, it also exhibits the lowest prediction accuracy and greatest sensitivity to measurement errors and unseen scenarios, as discussed in previous sections. In contrast, the proposed method achieves a compelling balance between computational efficiency and prediction accuracy, achieving an average inference time of $2.44$ s while exhibiting the highest accuracy. Furthermore, it outperforms the PF-Integrated ML method in both computation time and memory usage, demonstrating superior prediction accuracy, predictive robustness, and memory usage. This enhancement can be attributed to the adoption of the DL-ACPF formulation, which substantially reduces computational complexity compared to the PF-Integrated method.

\begin{table}[t]
\centering
\caption{\textsc{Computational Efficiency Comparison in the WECC 179-bus System}}
\label{tab:comput_eff}
\begin{tabular}{c c c}
\hline
\makecell{\#Method} & \makecell{Computation \\ Time [s]} & \makecell{Computation \\ Memory [MB]} \\ 
\hline
\textbf{Proposed}& \textbf{2.44} & \textbf{2068} \\
PF-Integrated ML \cite{IntegratingLiJiaming10066348} & 2.79 & 2571 \\
Data-driven LSTM & 0.63 & 1020 \\
 \textcolor{blue}{GNN}& \textcolor{blue}{3.14}&\textcolor{blue}{2944}\\
 \hline
\end{tabular}
\end{table}

\section{Conclusion}

This paper proposes a network interdependency-informed framework for online dynamic trajectory prediction in power systems with high IBR penetration. The proposed STAN-based predictor captures temporal dependencies in individual IBR bus dynamics and spatial interdependencies among neighboring buses, enabling accurate black-box modeling without access to proprietary control details. By incorporating the DL-ACPF equations into a hybrid physics-informed loss function, the model preserves system-wide physical consistency and enhances robustness against measurement errors. Case studies under various faults demonstrate that the proposed approach achieves superior accuracy, physical consistency, and robustness to measurement errors compared to recent ML methods. 

\textcolor{blue}{
The current phasor-domain formulation is intended for dynamic trajectories observable from PMU measurements.
As standard PMU phasor quantities do not fully capture the higher-frequency content associated with sub-synchronous oscillations and electromagnetic transient phenomena, such dynamics require EMT-domain modeling and measurements, which are left for further investigation.
Another future work will enhance the applicability of our method to incomplete or uncertain network information by integrating topology identification techniques.}

\appendix

LSTM layers are used in the proposed method to capture long-range temporal correlations in the voltage sequences. The LSTM adopts a memory cell structure that consists of the input gate $\mathbf{i}(t)$, the forget gate $\mathbf{f}(t)$, and the output gate $\mathbf{o}(t)$. These gates regulate the flow of information, which consists of the previous cell state $\mathbf{c}(t-1)$, the previous hidden state $\mathbf{H}(t-1)$, and the current input $\mathbf{u}(t) $, according to
\begin{align} \label{LSTM}
    \mathbf{f}(t) & = \sigma(\mathbf{u}(t)\mathbf{W}_{uf}+\mathbf{H}(t-1)\mathbf{W}_{hf}+\mathbf{b}_{f})\\
    \mathbf{i}(t) & = \sigma(\mathbf{u}(t)\mathbf{W}_{ui}+\mathbf{H}(t-1)\mathbf{W}_{hi}+\mathbf{b}_{i})\\
    \mathbf{o}(t) & = \sigma(\mathbf{u}(t)\mathbf{W}_{uo}+\mathbf{H}(t-1)\mathbf{W}_{ho}+\mathbf{b}_{o})
\end{align}
where $\mathbf{W}_{u\cdot}$ and $\mathbf{W}_{h\cdot}$ are input and recurrent weight matrices, and $\mathbf{b}_{\cdot}$ are the associated biases, corresponding to each gate; $\sigma(\cdot)$ denotes the element-wise sigmoid activation.

Subsequently, the candidate cell state $\hat{\mathbf{c}}(t)$, updated cell state $\mathbf{c}(t)$, and updated hidden state  $\mathbf{H}(t)$ are obtained by
\begin{align} 
    \hat{\mathbf{c}}(t) &= \tanh(\mathbf{u}(t)\mathbf{W}_{uc} + \mathbf{H}(t-1)\mathbf{W}_{hc}+\mathbf{b}_{c})\\ 
    \mathbf{c}(t) &= \mathbf{f}(t)\odot \mathbf{c}(t-1)+\mathbf{i}(t)\odot \hat{\mathbf{c}}(t) \label{LSTM_cell}
\end{align}
where $\odot$ denotes the element-wise multiplication; the cell state is updated by the forget gate, which determines the contribution of the previous cell state, and the input gate, which governs the influence of the current input.

Then, the hidden state is updated in \eqref{LSTM_hidden_2} by the output gate, which controls the contribution of the updated cell state to the hidden state.
\begin{align} 
    \mathbf{H}(t) &= \mathbf{o}(t)\odot \tanh(\mathbf{c}(t)) \label{LSTM_hidden_2}
\end{align}
\IEEEpubidadjcol
\bibliographystyle{IEEEtran}
\bibliography{IEEEabrv,Citation}
\let\mybibitem\bibitem

\end{document}